\documentclass[journal]{IEEEtran}
\ifCLASSINFOpdf
\else
\fi
\usepackage{booktabs}
\usepackage{multirow}
\usepackage{graphicx}
\usepackage{subfigure}
\usepackage{amssymb}
\usepackage{mathtools}
\usepackage{float}
\usepackage{color}
\usepackage{url} 
\usepackage{siunitx}
\begin{document}
%
\title{RDEIC: Accelerating Diffusion-Based Extreme Image Compression with Relay Residual Diffusion}
%
%
%
\author{Zhiyuan Li, Yanhui Zhou, Hao Wei, Chenyang Ge, Ajmal Mian
\thanks{This work are supported by the National Natural Science Foundation of China (NSFC62088102, NSFC62376208). (\emph{Corresponding author: Chenyang Ge.})}

\thanks{Zhiyuan Li, Yanhui Zhou, Hao Wei and Chenyang Ge are with the Institute of Artificial Intelligence and Robotics, Xi'an Jiaotong University, Xi'an 710049, China (e-mail: lizhiyuan2839@163.com; zhouyh@mail.xjtu.edu.cn; haowei@stu.xjtu.edu.cn; cyge@mail.xjtu.edu.cn)}

\thanks{Ajmal Mian is with the Department of Computer Science and Software Engineering, The University of Western Australia, Perth, Crawley, WA 6009, Australia (e-mail: ajmal.mian@uwa.edu.au)}
}
\maketitle

\begin{abstract}
Diffusion-based extreme image compression methods have achieved impressive performance at extremely low bitrates. However, constrained by the iterative denoising process that starts from pure noise, these methods are limited in both fidelity and efficiency. To address these two issues, we present Relay Residual Diffusion Extreme Image Compression (\textbf{RDEIC}), which leverages compressed feature initialization and residual diffusion. Specifically, we first use the compressed latent features of the image with added noise, instead of pure noise, as the starting point to eliminate the unnecessary initial stages of the denoising process. Second, we directly derive a novel residual diffusion equation from Stable Diffusion's original diffusion equation that reconstructs the raw image by iteratively removing the added noise and the residual between the compressed and target latent features. In this way, we effectively combine the efficiency of residual diffusion with the powerful generative capability of Stable Diffusion. Third, we propose a fixed-step fine-tuning strategy to eliminate the discrepancy between the training and inference phases, thereby further improving the reconstruction quality. Extensive experiments demonstrate that the proposed RDEIC achieves state-of-the-art visual quality and outperforms existing diffusion-based extreme image compression methods in both fidelity and efficiency. The source code and pre-trained models are available at \url{https://github.com/huai-chang/RDEIC}.
\end{abstract}

\section{Introduction}
Extreme image compression is becoming increasingly important with the growing demand for efficient storage and transmission of images where storage capacity or bandwidth is limited, such as in satellite communications and mobile devices. Conventional compression standards like JPEG \cite{JPEG}, BPG \cite{BPG} and VVC \cite{VVC} rely on hand-crafted rules and block-based redundancy removal techniques, leading to severe blurring and blocking artifacts at low bitrates. Hence, there is an urgent need to explore extreme image compression methods.

\begin{figure}[t]\scriptsize
\centering
\includegraphics[width=0.48\textwidth]{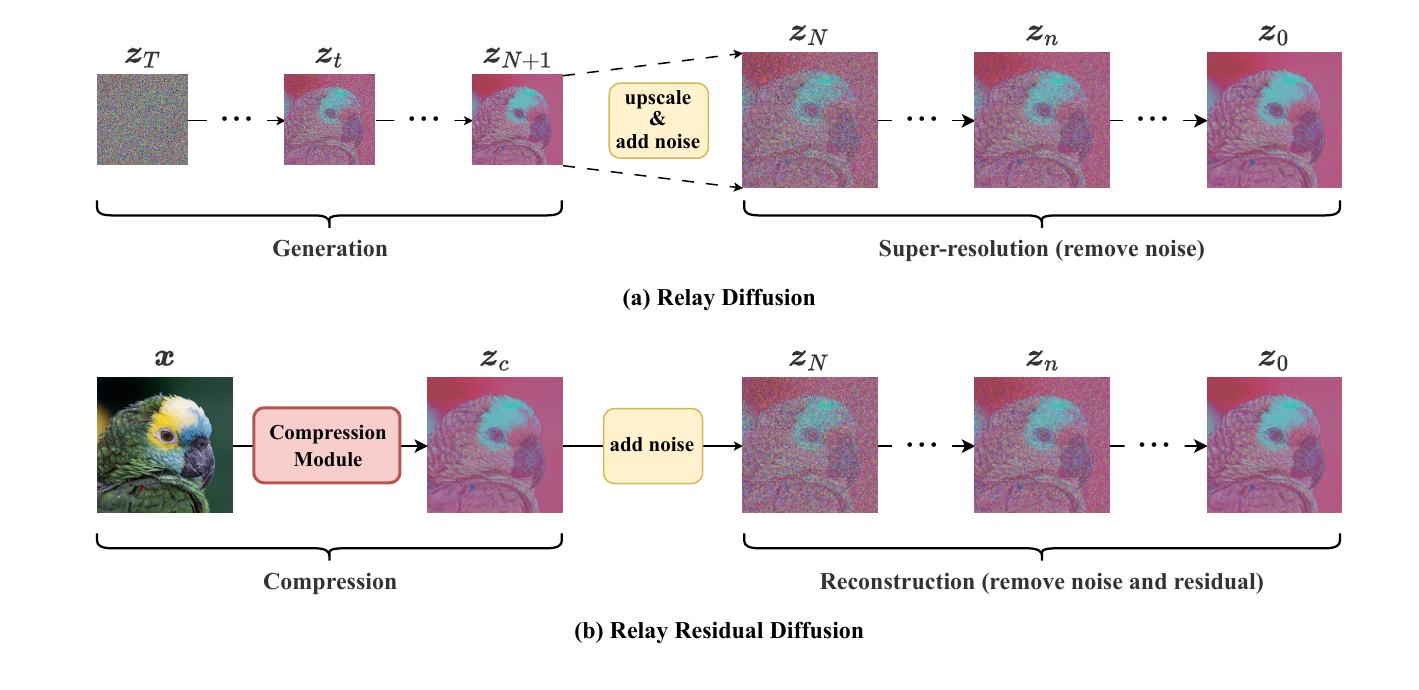}
\caption{{The architecture of relay diffusion and our relay residual diffusion. Relay diffusion consists of two sequential stages. In the first stage, a low-resolution latent (or image) $\boldsymbol{z}_{N+1}$ is generated from pure noise $\boldsymbol{z}_T$ through a standard diffusion process. This intermediate result is then upsampled and injected with noise to produce $\boldsymbol{z}_N$, which serves as the starting point for the super-resolution stage. This design enables the model to preserve structural consistency across resolutions while avoiding redundant sampling from pure noise in later stages, improving both quality and efficiency. Inspired by relay diffusion, our relay residual diffusion reinterpret image compression and reconstruction as an analogous pair. We replace the generation stage with a compression stage, where the input image is encoded into compressed latent features $\boldsymbol{z}_c$. We then inject noise into $\boldsymbol{z}_c$ to obtain the starting point $\boldsymbol{z}_N$ for reconstruction stage. Since $\boldsymbol{z}_N$ contains both injected noise and the residual between $\boldsymbol{z}_c$ and the target latent $\boldsymbol{z}_0$, we propose a novel residual diffusion process that jointly removes both components.
}} 
\label{RD&RRD}
\end{figure}

In recent years, learned image compression methods have attracted significant interest, outperforming conventional codecs. However, distortion-oriented learned compression methods \cite{INC, ViTLIC, TCM, FAT} optimize for the rate-distortion function alone, resulting in unrealistic reconstructions at low bitrates, typically manifested as blurring or over-smoothing. Perceptual-oriented learned compression methods \cite{GANELIC, HiFiC, MS-ILLM, CDC} introduce generative models, such as generative adversarial networks (GANs) \cite{GAN} and diffusion models \cite{DDPM}, to enhance the perceptual quality of reconstructions. However, these methods are optimized for medium to high bitrates instead of extremely low bitrates such as below 0.1 bpp. As a result, these methods experience significant quality degradation when the compression ratio is increased.

\begin{figure*}[t]\scriptsize
\centering
\makebox[0.135\textwidth]{\textbf{Original}}
\makebox[0.135\textwidth]{\textbf{VVC}}
\makebox[0.135\textwidth]{\textbf{MS-ILLM}}
\makebox[0.135\textwidth]{\textbf{Text+Sketch-25}}
\makebox[0.135\textwidth]{\textbf{PerCo-20}}
\makebox[0.135\textwidth]{\textbf{DiffEIC-50}}
\makebox[0.135\textwidth]{\textbf{RDEIC-5 (Ours)}}
\\ \vspace{0.1cm}
\includegraphics[width=0.135\textwidth]{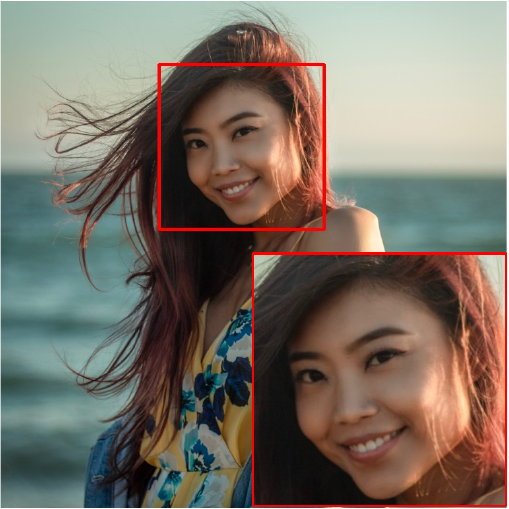}
\includegraphics[width=0.135\textwidth]{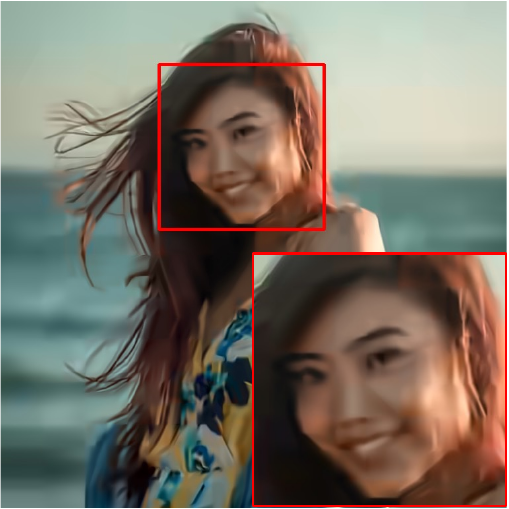}
\includegraphics[width=0.135\textwidth]{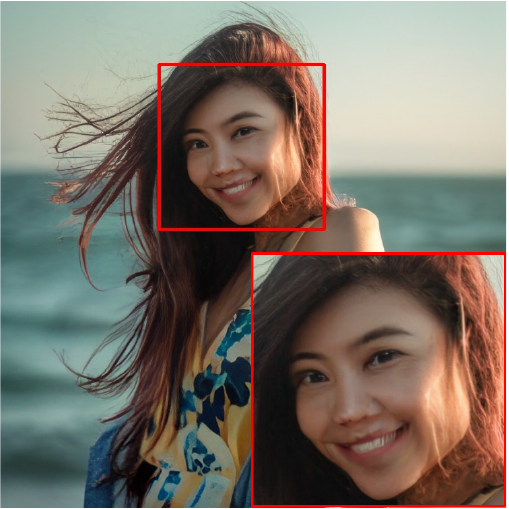}
\includegraphics[width=0.135\textwidth]{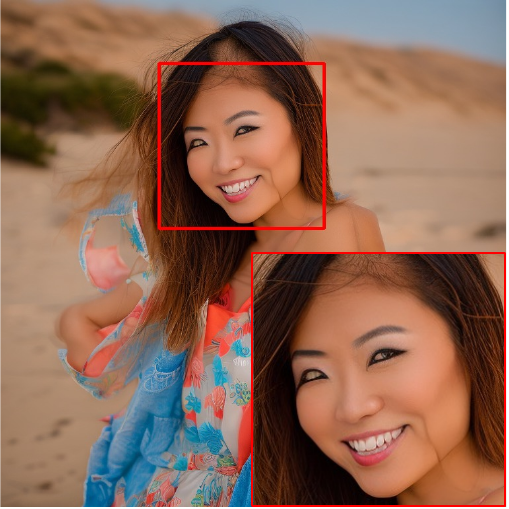}
\includegraphics[width=0.135\textwidth]{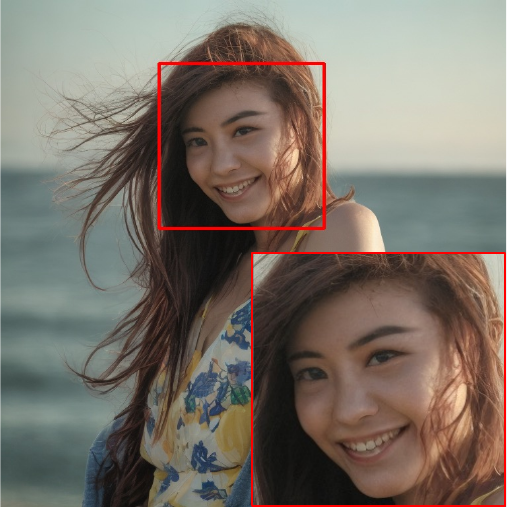}
\includegraphics[width=0.135\textwidth]{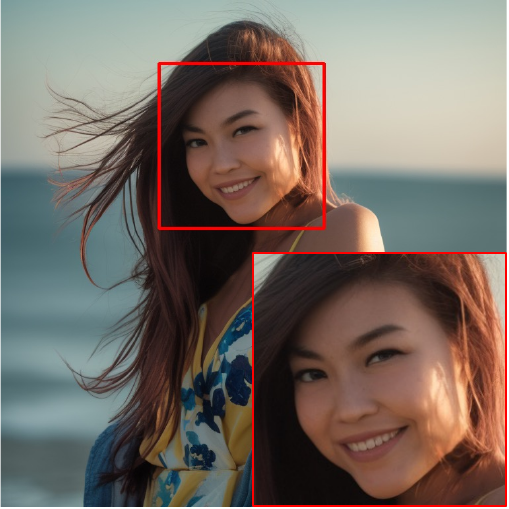}
\includegraphics[width=0.135\textwidth]{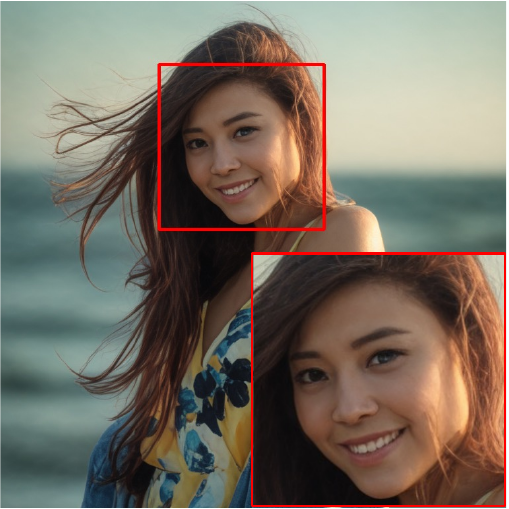}
\\ \vspace{0.1cm}
\makebox[0.135\textwidth]{\textbf{bpp / DISTS$\downarrow$}}
\makebox[0.135\textwidth]{\textbf{0.0340 / 0.2029}}
\makebox[0.135\textwidth]{\textbf{0.0327 / 0.0867}}
\makebox[0.135\textwidth]{\textbf{0.0237 / 0.1959}}
\makebox[0.135\textwidth]{\textbf{0.0320 / 0.1177}}
\makebox[0.135\textwidth]{\textbf{0.0278 / 0.1019}}
\makebox[0.135\textwidth]{\textbf{0.0199 / 0.0989}}
\\ \vspace{0.1cm}
\includegraphics[width=0.135\textwidth]{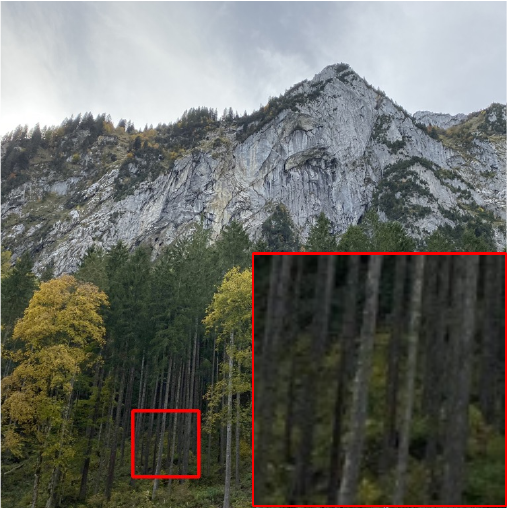}
\includegraphics[width=0.135\textwidth]{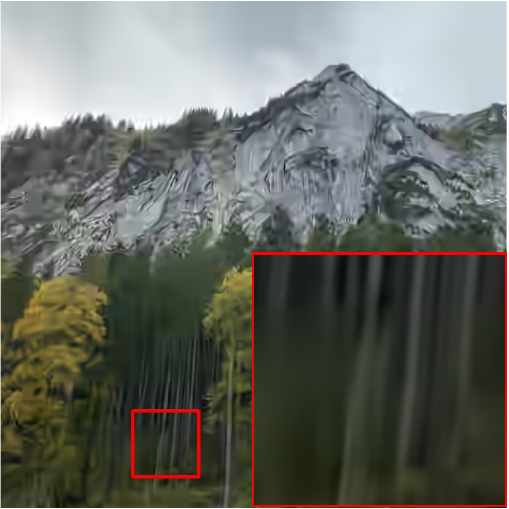}
\includegraphics[width=0.135\textwidth]{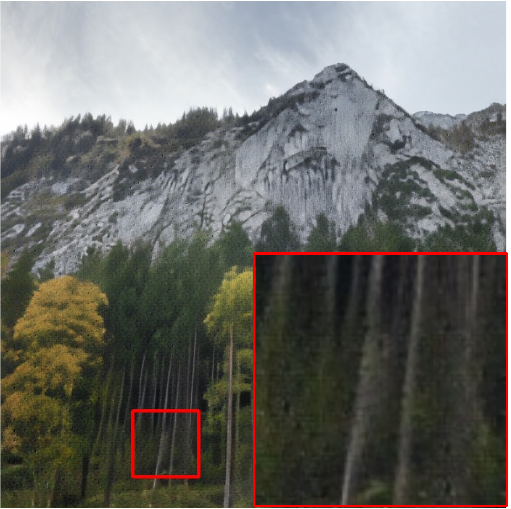}
\includegraphics[width=0.135\textwidth]{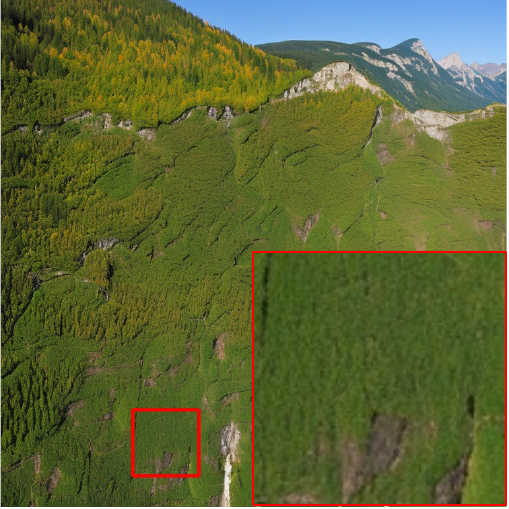}
\includegraphics[width=0.135\textwidth]{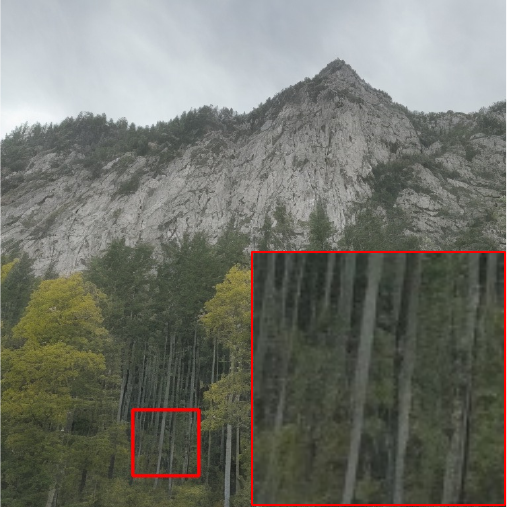}
\includegraphics[width=0.135\textwidth]{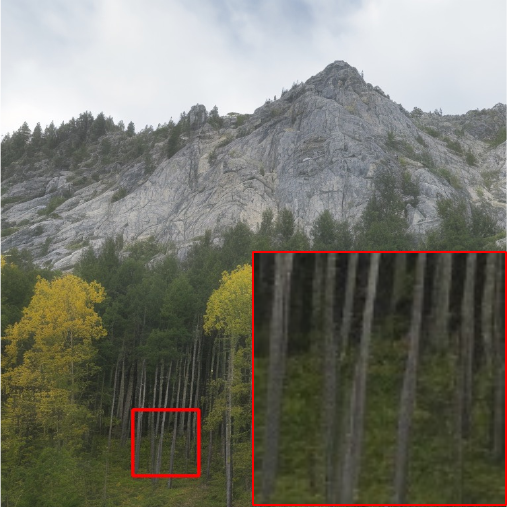}
\includegraphics[width=0.135\textwidth]{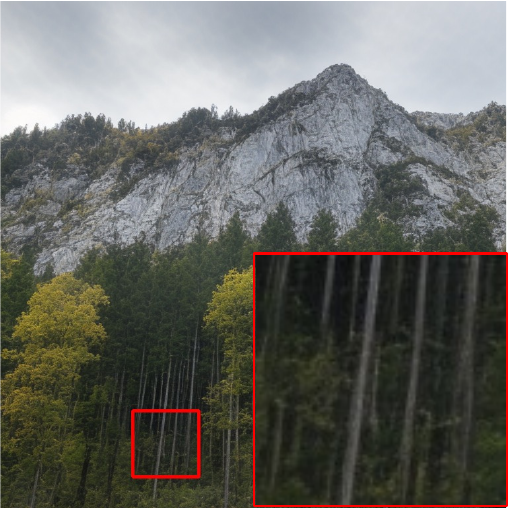}
\\ \vspace{0.1cm}
\makebox[0.135\textwidth]{\textbf{bpp / DISTS$\downarrow$}}
\makebox[0.135\textwidth]{\textbf{0.0461 / 0.3293}}
\makebox[0.135\textwidth]{\textbf{0.0357 / 0.1986}}
\makebox[0.135\textwidth]{\textbf{0.0236 / 0.2783}}
\makebox[0.135\textwidth]{\textbf{0.0321 / 0.2256}}
\makebox[0.135\textwidth]{\textbf{0.0176 / 0.1812}}
\makebox[0.135\textwidth]{\textbf{0.0176 / 0.1700}}
\\ \vspace{0.1cm}
\includegraphics[width=0.135\textwidth]{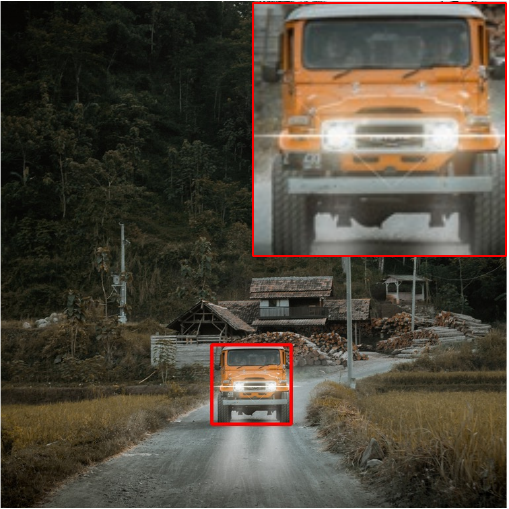}
\includegraphics[width=0.135\textwidth]{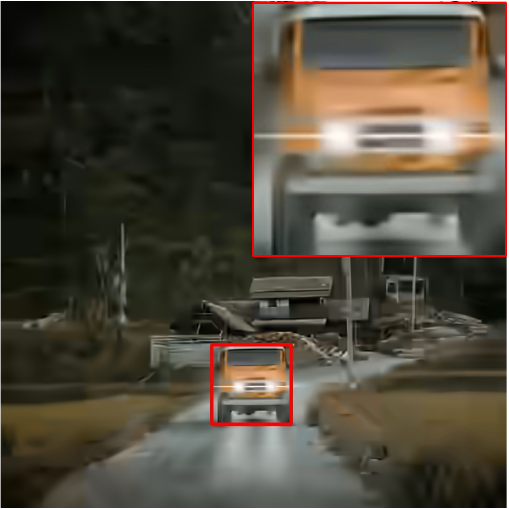}
\includegraphics[width=0.135\textwidth]{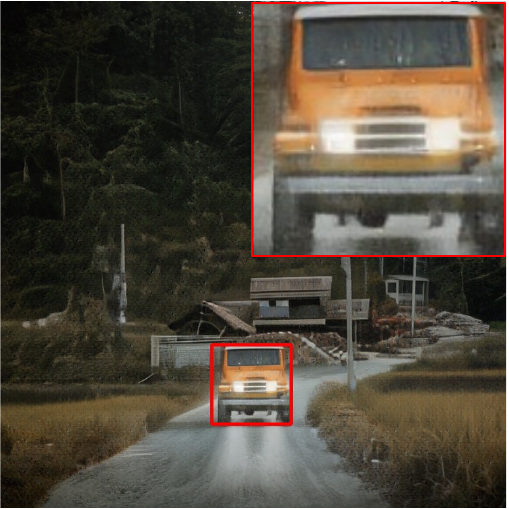}
\includegraphics[width=0.135\textwidth]{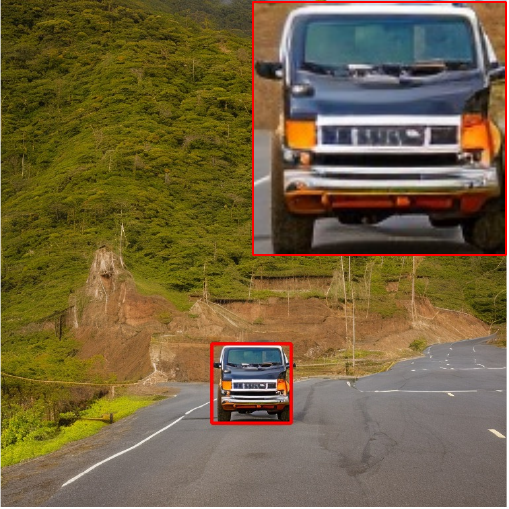}
\includegraphics[width=0.135\textwidth]{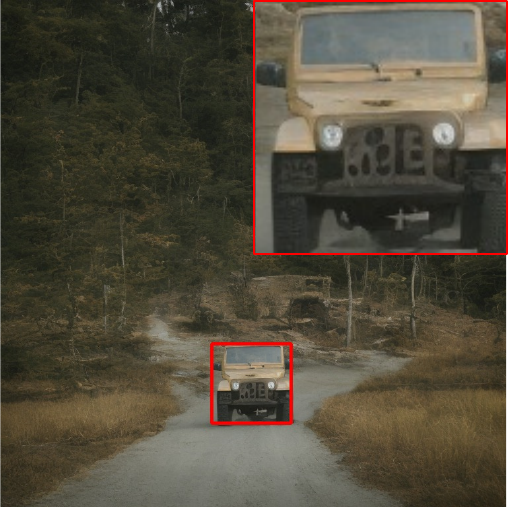}
\includegraphics[width=0.135\textwidth]{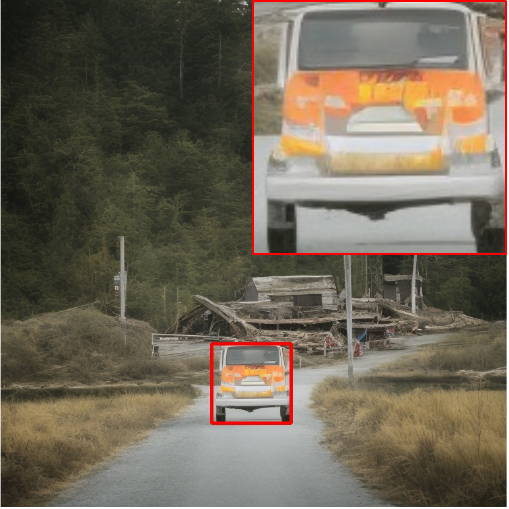}
\includegraphics[width=0.135\textwidth]{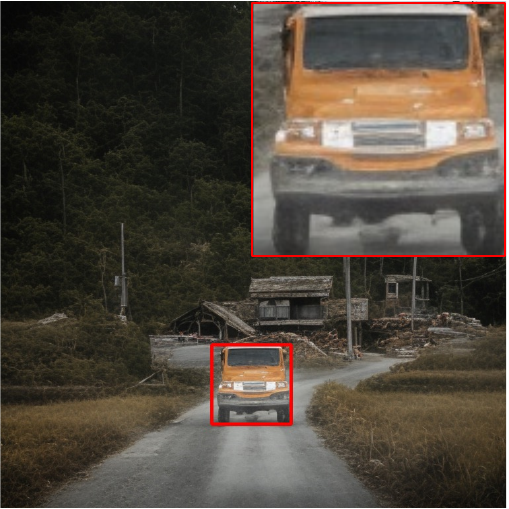}
\\ \vspace{0.1cm}
\makebox[0.135\textwidth]{\textbf{bpp / DISTS$\downarrow$}}
\makebox[0.135\textwidth]{\textbf{0.0284 / 0.3550}}
\makebox[0.135\textwidth]{\textbf{0.0324 / 0.1727}}
\makebox[0.135\textwidth]{\textbf{0.0197 / 0.2680}}
\makebox[0.135\textwidth]{\textbf{0.0321 / 0.2595}}
\makebox[0.135\textwidth]{\textbf{0.0196 / 0.2105}}
\makebox[0.135\textwidth]{\textbf{0.0179 / 0.1752}}
\caption{Qualitative comparisons between the proposed RDEIC and state-of-the-art methods. The number of denoising steps is written
after the name, e.g. DiffEIC-50 means 50 diffusion steps are used by DiffEIC. The bpp and DISTS of each method are shown at the bottom of each image.}
\label{exhibition}
\end{figure*}

Recently, diffusion-based extreme image compression methods \cite{Text+Sketch, PerCo, DiffEIC} leverage the robust generative ability of pre-trained text-to-image (T2I) diffusion models, achieving superior visual quality at extremely low bitrates. Nonetheless, these methods are constrained by the inherent characteristics of diffusion models. Firstly, these methods rely on an iterative denoising process to reconstruct raw images from pure noise, which is inefficient for inference \cite{DiffEIC}. Secondly, initiating the denoising process from pure noise introduces significant randomness, compromising the fidelity of the reconstructions \cite{PerCo}. Thirdly, there is a discrepancy between the training and inference phases. During training, each time-step is trained independently, which is well-suited for image generation tasks where diversity (or randomness) is encouraged \cite{DDPM}. However, this training approach is not optimal for image compression where consistency between the reconstruction and the raw image is crucial.

In this work, we propose \textbf{R}elay \textbf{R}esidual \textbf{D}iffusion \textbf{E}xtreme \textbf{I}mage \textbf{C}ompression (\textbf{RDEIC}) to overcome the three limitations mentioned above. To overcome the first two limitations, we proposed a novel relay residual diffusion framework, {as shown in Fig. \ref{RD&RRD}(b)}. Specifically, we construct the starting point using the compressed latent features combined with slight noise, transitioning between the starting point and target latent features by shifting the residual between them. This approach significantly reduces the number of denoising steps required for reconstruction while ensures that the starting point retains most of the information from the compressed features, providing a strong foundation for subsequent detail generation. To leverage the robust generative capability of pre-trained stable diffusion for extreme image compression, we derive a novel residual diffusion equation directly from stable diffusion's diffusion equation, rather than designing a new diffusion equation from scratch as done in \cite{ResShift}. To address the third limitation, we introduce a fixed-step fine-tuning strategy to eliminate the discrepancy between the training and inference phases. By fine-tuning RDEIC throughout the entire reconstruction process, we further improve the reconstruction quality. Moreover, to meet users’ diverse requirements, we introduce a controllable detail generation method that achieves a trade-off between smoothness and sharpness by adjusting the intensity of high-frequency components in the reconstructions. As shown in Fig. \ref{exhibition}, the proposed RDEIC achieves state-of-the-art perceptual performance at extremely low bitrates, and significantly outperforms existing diffusion-based extreme image compression methods with fewer inference steps.

In summary, the main contributions of this work are as follows:

    
1) We propose a {\textbf{R}elay \textbf{R}esidual \textbf{D}iffusion (\textbf{RRD})} process that effectively combines the efficiency of residual diffusion with the powerful generative capability of stable diffusion. To the best of our knowledge, we are the first to successfully integrate stable diffusion into a residual diffusion framework.
    
2) To eliminate the discrepancy between the training and inference phases, we design a {\textbf{F}ixed-\textbf{S}tep \textbf{F}ine-\textbf{T}uning (\textbf{FSFT})} strategy that refines the model through the entire reconstruction process, further improving reconstruction quality.

3) We introduce a controllable detail generation method to balance smoothness and sharpness, allowing users to explore and customize outputs according to their personal preferences.

The remainder of this paper is organized as follows. The related works are summarized in Section \ref{related_work}. The proposed method is described in Section \ref{proposed_method}. The experiment results and analysis are presented in Section \ref{experiments} and Section \ref{analysis}, respectively. Finally, we conclude our work in Section \ref{conclusion}. 
\section{Related Work}
\label{related_work}
\subsection{Learned Image Compression}
As a pioneer work, Ball{\'e} et al. \cite{E2EOIC} proposed an VAE-based image compression framework to jointly optimize the rate-distortion performance. In \cite{Hyperprior}, they later introduced a hyperprior to reduce spatial dependencies in the latent representation, greatly enhancing performance. Subsequent works further improved compression models by developing various nonlinear transforms \cite{INC, TCM, FAT, ELIC} and entropy models \cite{Autoregressive, Channel-wise, Checkerboard, Entroformer, CCP}. However, optimization for rate-distortion alone often results in unrealistic reconstructions at low bitrates, typically manifested as blurring or over-smoothness \cite{RDP-tradeoff}.

To improve perceptual quality, generative models have been integrated into compression methods. Agustsson et al. \cite{GANELIC} added an adversarial loss for lost details generation. Mentzer et al. \cite{HiFiC} explored the generator and discriminator architectures, as well as training strategies for perceptual image compression. In \cite{MS-ILLM}, Muckley et al. introduced a local adversarial discriminator to enhance statistical fidelity. With the advancement of diffusion models, some efforts have been made to apply diffusion models to image compression. For instance, Yang et al. \cite{CDC} innovatively introduced a conditional diffusion model as decoder for image compression. Kuang et al. \cite{CGDM} proposed a consistency guidance architecture to guide the diffusion model in stably reconstructing high-quality images. {In addition to natural image compression, learned methods have been extended to specialized image modalities. For example, stereo image compression jointly encodes left-right image pairs by exploiting inter-view redundancy, achieving higher compression efficiency compared to independent coding \cite{HESIC, MASIC}. Meanwhile, satellite image compression focuses on efficiently handling ultra-high-resolution aerial or satellite imagery with large spatial dimensions and complex textures \cite{cosmic}.}

{In our work, we adopt a VAE-based compression backbone to effectively extract and compress image information, and integrate a diffusion-based reconstruction module on the decoder side to improve reconstruction quality at extremely low bitrates.}

\subsection{Extreme Image Compression}
In recent years, extreme image compression has garnered increasing attention, aiming to compress image to extremely low bitrates, often below 0.1 bpp, while maintaining visually acceptable image quality. Gao et al. \cite{INVEIC} leveraged the information-lossless property of invertible neural networks to mitigate the significant information loss in extreme image compression. Jiang et al. \cite{TGIC} treated text descriptions as prior to ensure semantic consistency between the reconstructions and the raw images. In \cite{VQIR}, Wei et al. achieved extreme image compression by rescaling images using extreme scaling factors. Lu et al. \cite{hybridflow} combined continuous and codebook-based discrete features to reconstruct high-quality images at extremely low bitrates. 

Inspired by the great success of T2I diffusion models in various image restoration tasks \cite{DiffBIR, StableSR}, some methods have incorporated T2I diffusion models into extreme image compression frameworks. Lei et al. \cite{Text+Sketch} utilized a pre-trained ControlNet \cite{ControlNet} to reconstruct images based on corresponding short text prompts and binary contour sketches. Careil et al. \cite{PerCo} conditioned iterative diffusion models on vector-quantized latent image representations and textual image descriptions. In our prior work \cite{DiffEIC}, we combined compressive VAEs with pre-trained T2I diffusion models to achieve realistic reconstructions at extremely low bitrates. 

However, constrained by the inherent characteristics of diffusion models, these diffusion-based extreme image compression methods are limited in both fidelity and efficiency. In this paper, we propose a solution to these limitations through a relay residual diffusion framework and a fixed-step fine-tuning strategy.

\subsection{Relay Diffusion}

To achieve high-resolution image generation, cascaded diffusion methods \cite{Cascaded, photorealistic} decompose the image generation into multiple stages, with each stage responsible for super-resolution conditioning on the previous one. However, these methods still require complete resampling at each stage, leading to inefficiencies and potential mismatches among different resolutions. 

Relay diffusion, as proposed by Teng et al. \cite{relaydiffusion}, extends the cascaded diffusion by continuing the diffusion process directly from the low-resolution output rather than restarting from pure noise, which allows the higher-resolution stages to correct artifacts from earlier stages, as shown in Fig. \ref{RD&RRD}(a). This design is particularly well-suited for tasks such as image restoration and image compression, where degraded images or features are available. For instance, PASD \cite{PASD} and SeeSR \cite{SeeSR} directly embed the LR latent into the initial random noise during the inference process to alleviate the inconsistency between training and inference. ResShift \cite{ResShift} further constructs a Markov chain that transfers between degraded and target features by shifting the residual between them, substantially improving the transition efficiency. However, its redesigned diffusion equation and noise schedule prevent it from leveraging the robust generative capability of pre-trained stable diffusion. 

In this work, we directly derive a new relay residual diffusion equation from stable diffusion's diffusion equation, effectively combining the efficiency of residual diffusion with the powerful generative capability of stable diffusion.

\section{Methodology}
\label{proposed_method}

\begin{figure*}[t]
\begin{center}
\includegraphics[width=0.98\textwidth]{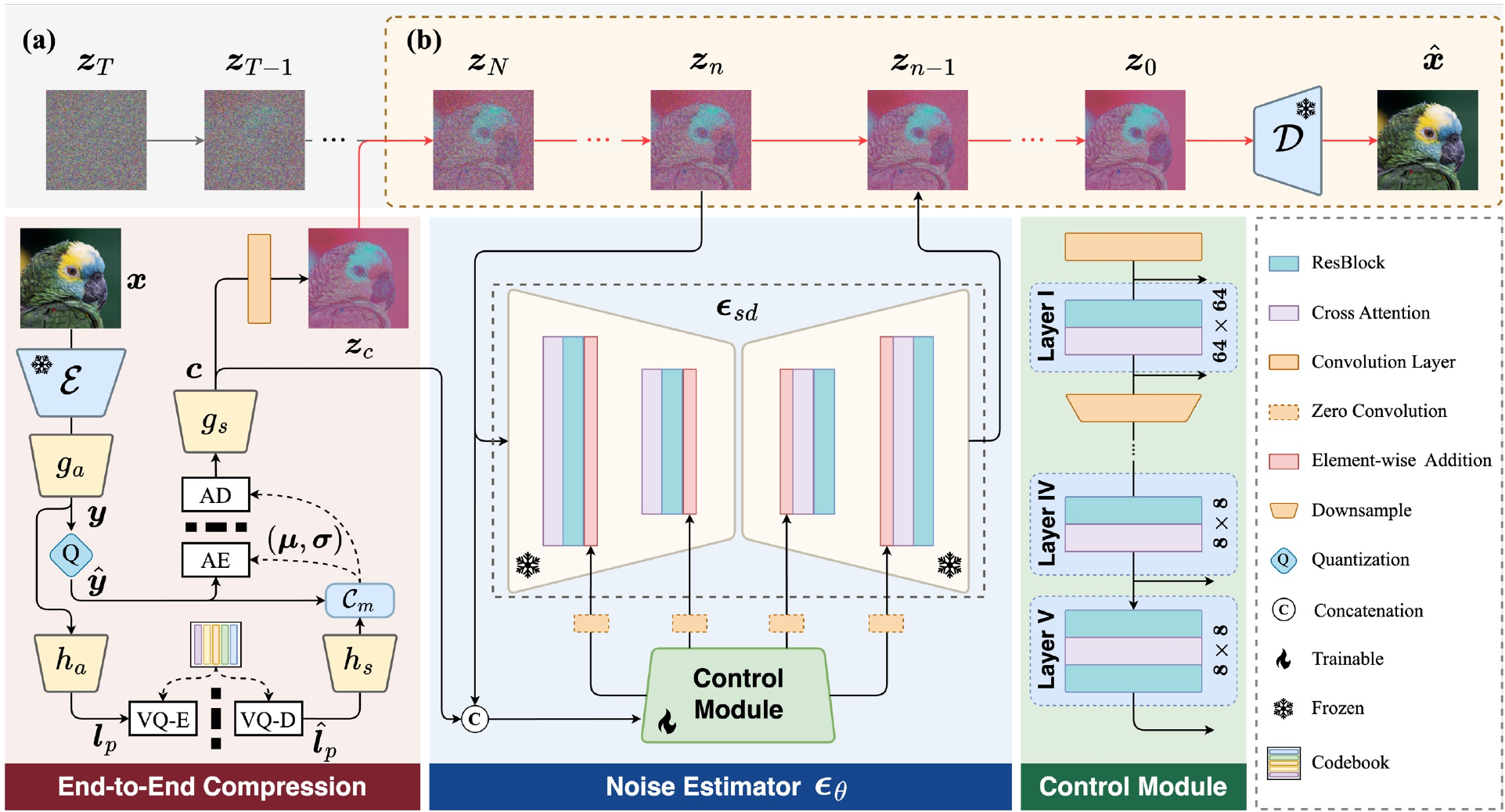}
\end{center}
\caption{The proposed RDEIC. We first map a raw image $\boldsymbol{x}$ into the latent space using the encoder $\mathcal{E}$ and then perform end-to-end lossy compression to get {intermediate representations $\boldsymbol{c}$ (with 256 channels)} and compressed latent features $\boldsymbol{z}_c$. We then use $\boldsymbol{z}_c$ with added noise as the starting point and apply a denoising process to reconstruct the noise-free latent feature $\boldsymbol{z}_0$. {Note that, since $\boldsymbol{c}$ can carry more information, we choose it—rather than the more compact $z_c$—as the input to the control module.} The decoder $\mathcal{D}$ maps $\boldsymbol{z}_0$ back to the pixel space, to get the reconstructed image $\hat{\boldsymbol{x}}$. (a) Vanilla diffusion framework that starts from pure noise. (b) The proposed relay residual diffusion framework that starts from compressed latent features with added noise.}
\label{RDEIC}
\end{figure*}

\subsection{Overall Framework}
Fig. \ref{RDEIC} shows an overview of the proposed RDEIC network. 
We first use an encoder $\mathcal{E}$ and analysis transform $g_a$ to convert the input image $\boldsymbol{x}$ to its latent representation $\boldsymbol{y}$. Then we perform hyper transform coding on $\boldsymbol{y}$ with the categorical hyper model \cite{GLC} and use the space-channel context model $\mathcal{C}_m$ to predict the entropy parameters $(\boldsymbol{\mu}, \boldsymbol{\sigma})$ to estimate the distribution of quantized latent representation $\boldsymbol{\hat{y}}$ \cite{ELIC}. The side information $\boldsymbol{l}_p$ is quantized through vector-quantization, i.e., $\hat{\boldsymbol{l}}_p$ is the mapping of $\boldsymbol{l}_p$ to its closest codebook entry. Subsequently, the synthesis transform $g_s$ is used to obtain { the intermediate representation $\boldsymbol{c}$ (with 256 channels) and the compressed features $\boldsymbol{z}_c$ (with 4 channels).} Random noise is then added to $\boldsymbol{z}_c$, which is the starting point for reconstructing the noise-free latent features $\boldsymbol{z}_0$ through an iterative denoising process. The denoising process is implemented by a frozen pre-trained noise estimator $\epsilon_{sd}$ of stable diffusion with trainable control module for intermediate feature modulation. {Note that, since $\boldsymbol{c}$ can carry more information, we choose it, rather than the more compact $\boldsymbol{z}_c$, as the input to control module.} Finally, the reconstructed image $\boldsymbol{\hat{x}}$ is decoded from  $\boldsymbol{z}_0$ using the decoder $\mathcal{D}$.

\subsection{Relay Residual Diffusion}
Following stable diffusion, existing diffusion-based extreme image compression methods obtain the noisy latent by adding Gaussian noise with variance $\beta_t \in (0,1)$ to the noise-free latent features $\boldsymbol{z}_0$:
\begin{equation}
    \boldsymbol{z}_t = \sqrt{\bar{\alpha}_{t}}\boldsymbol{z}_0 + \sqrt{1-\bar{\alpha}_{t}} \epsilon_t, \ t=1,2,\cdots,T,
    \label{z_t}
\end{equation}
where $\epsilon_t \sim \mathcal{N}(0, \boldsymbol{I})$, $\alpha_t = 1-\beta_t$ and $\bar{\alpha}_t=\prod_{i=1}^t \alpha_i$. When $t$ is large enough, the noisy latent $\boldsymbol{z}_t$ is nearly a standard Gaussian distribution. In practice, $T$ is typically very large, e.g., 1000, and pure noise is set as the starting point for the reverse diffusion process. {The reverse process at time step t is defined as:
\begin{equation}
    z_{t-1} = \underbrace{\frac{\sqrt{\bar{\alpha}_{t-1}}\beta_t}{1-\bar{\alpha}_t}z_0 + \frac{\sqrt{\bar{\alpha}_t}(1-\bar{\alpha}_{t-1}))}{1-\bar{\alpha}_t}z_t}_{\mu_t(z_t,\ z_0)} + \underbrace{\sqrt{\frac{1-\bar{\alpha}_{t-1}}{1-\bar{\alpha}_{t}}\beta_t}}_{\sigma_t}\epsilon,
    \label{ddpm}
\end{equation}
Where $\epsilon \sim \mathcal{N}(0,I)$.} However, this approach is not optimal for the image compression task, where the compressed latent features $\boldsymbol{z}_c$ are available.

\begin{figure*}[t]\scriptsize
\begin{center}
\includegraphics[width=0.9\textwidth]{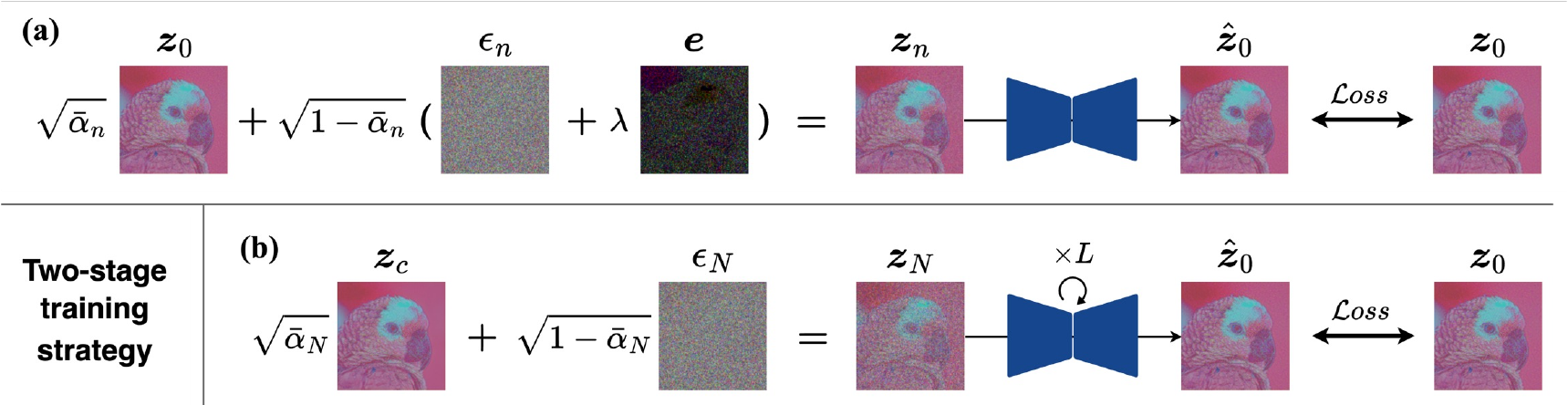}
\end{center}
\caption{The two-stage training strategy of RDEIC. (a) Independent training: we randomly pick a time-step $n$ and train each time-step $n$ independently. This ensures that the model effectively learns to remove added noise and residuals at every step.  (b) Fixed-step fine-tuning: $L$ fixed denoising steps are used to iteratively reconstruct a noise-free latent features $\hat{\boldsymbol{z}}_0$ from $\boldsymbol{z}_N$, which is consistent with the inference phase.} 
\label{FSFT}
\end{figure*}

To this end, we set the starting point to
$\boldsymbol{z}_N = \sqrt{\bar{\alpha}_N} \boldsymbol{z}_c + \sqrt{1-\bar{\alpha}_N}\epsilon_N$, where $N \ll T$. Our relay residual diffusion is thus defined as:
\begin{equation}
    \boldsymbol{z}_n = \sqrt{\bar{\alpha}_{n}}(\boldsymbol{z}_0 + \eta_n \boldsymbol{e})+ \sqrt{1-\bar{\alpha}_{n}} \epsilon_n, \ n=1,2,\cdots,N,
    \label{z_n}
\end{equation}
where $\boldsymbol{e}$ denotes the residual between $\boldsymbol{z}_c$ and $\boldsymbol{z}_0$, i.e., $\boldsymbol{e}=\boldsymbol{z}_c-\boldsymbol{z}_0$, and $\{ \eta_n \}_{n=1}^{N}$ is a weight sequence that satisfies $\eta_1 \to 0$ and $\eta_N=1$. Moreover, according to Eq. (\ref{z_n}), $\boldsymbol{z}_{n-1}$ can be sampled as:
\begin{align}
    \boldsymbol{z}_{n-1} &= \sqrt{\bar{\alpha}_{n-1}}(\boldsymbol{z}_0 + \eta_{n-1} \boldsymbol{e})+ \sqrt{1-\bar{\alpha}_{n-1}} \epsilon_{n-1}, \\
    &= \sqrt{\bar{\alpha}_{n-1}}\boldsymbol{z}_0 + \sqrt{\bar{\alpha}_{n-1}} \eta_{n-1} \boldsymbol{e} + \underbrace{\sqrt{1-\bar{\alpha}_{n-1}} \epsilon_{n-1}}_{\sim \mathcal{N}(0, (1-\bar{\alpha}_{n-1})\boldsymbol{I})}.
    \label{z_n-1.1}
\end{align}
Since the residual $\boldsymbol{e}$ is unavailable during inference, {we refer to Eq. (\ref{ddpm}) and define $\boldsymbol{z}_{n-1}$ as a linear combination of $\boldsymbol{z}_n$ and $\boldsymbol{z}_0$:
\begin{equation}
    \boldsymbol{z}_{n-1} = \underbrace{k_n \boldsymbol{z}_{0} + m_n \boldsymbol{z}_{n}}_{\mu_n(\boldsymbol{z}_n,\ \boldsymbol{z}_0)} + \sigma_n \epsilon,
    \label{z_n-1}
\end{equation}}
Substituting $\boldsymbol{z}_n$, as defined in Eq. (\ref{z_n}), into Eq. (\ref{z_n-1}), we obtain:
\begin{equation}
    \begin{split}
        \boldsymbol{z}_{n-1} = (k_n + m_n \sqrt{\bar{\alpha}_{n}}) \boldsymbol{z}_0 + m_n \sqrt{\bar{\alpha}_{n}} \eta_n \boldsymbol{e} + \\ \underbrace{m_n \sqrt{1-\bar{\alpha}_{n}} \epsilon_n + \sigma_n \epsilon}_{\sim \mathcal{N}(0, (m_n^2 (1-\bar{\alpha}_{n}) + \sigma_n^2)\boldsymbol{I})}.
    \end{split}
    \label{z_n-1.2}
\end{equation}
By combining Eq. (\ref{z_n-1.1}) and Eq. (\ref{z_n-1.2}), we obtain the following equations:
\begin{equation}
\begin{cases}
    \sqrt{\bar{\alpha}_{n-1}} = k_n + m_n \sqrt{\bar{\alpha}_{n}}, \\
    \sqrt{\bar{\alpha}_{n-1}} \eta_{n-1} = m_n \sqrt{\bar{\alpha}_{n}} \eta_n, \\
    1-\bar{\alpha}_{n-1} = m_n^2 (1-\bar{\alpha}_{n}) + \sigma_n^2.
\end{cases}
\label{A.1.3}
\end{equation}
Note that, referring to DDIM \cite{DDIM}, we set $\sigma_n = 0$ for simplicity. By solving Eq. (\ref{A.1.3}), we have:
\begin{equation}
    \frac{\eta_n}{\eta_{n-1}} = \frac{\sqrt{1-\bar{\alpha}_n}/\sqrt{\bar{\alpha}_{n}}}{\sqrt{1-\bar{\alpha}_{n-1}}/\sqrt{\bar{\alpha}_{n-1}}} \to \eta_n = \lambda \frac{\sqrt{1-\bar{\alpha}_n}}{\sqrt{\bar{\alpha}_{n}}},
    \label{eta}
\end{equation}
where we set $\lambda=\frac{\sqrt{\bar{\alpha}_N}}{\sqrt{1-\bar{\alpha}_N}}$ to ensure $\eta_N=1$. Substituting Eq. (\ref{eta}) into Eq. (\ref{z_n}), the diffusion process can be further written as follows:
\begin{align}
    \boldsymbol{z}_{n} &= \sqrt{\bar{\alpha}_n} (\boldsymbol{z}_0 + \lambda \frac{\sqrt{1-\bar{\alpha}_n}}{\sqrt{\bar{\alpha}_{n}}} \boldsymbol{e}) + \sqrt{1-\bar{\alpha}_n} \epsilon_n, \\
    &= \sqrt{\bar{\alpha}_n} \boldsymbol{z}_0 + \sqrt{1-\bar{\alpha}_n} \underbrace{(\lambda \boldsymbol{e} + \epsilon_n)}_{\tilde{\epsilon}_n}.
    \label{z_n1}
\end{align}
It is evident that Eq. (\ref{z_n1}) has the same structure as Eq. (\ref{z_t}), allowing us to seamlessly incorporate stable diffusion into our framework. For the denoising process, the noise estimator $\epsilon_\theta$ is learned to predict $\tilde{\epsilon}_n$ at each time-step $n$. The optimization of noise estimator $\epsilon_\theta$ is defined as
\begin{align}
    \mathcal{L}_{ne} &= \mathbb{E}_{\boldsymbol{z}_0,\boldsymbol{z}_c,\boldsymbol{c},n,\epsilon_n}\Vert \boldsymbol{z}_0 - \hat{\boldsymbol{z}}_0\Vert^2_2 \label{l_rd}, \\
    &=\omega_n \mathbb{E}_{\boldsymbol{z}_0,\boldsymbol{z}_c,\boldsymbol{c},n,\epsilon_n}\Vert \tilde{\epsilon}_n - \epsilon_\theta(\boldsymbol{z}_n, \boldsymbol{c}, n) \Vert^2_2,
    \label{l_rd1}
\end{align}
where $\omega_n=\frac{1-\bar{\alpha}_n}{\bar{\alpha}_n}$. After that, we can start from the compressed latent features $\boldsymbol{z}_c$ and reconstruct the image using Eq. (\ref{z_n-1}) without knowing the residual $\boldsymbol{e}$.

\subsection{Fixed-Step Fine-Tuning Strategy}
\label{FSFTS}
Most existing diffusion-based image compression methods adopt the same training strategy as DDPM \cite{DDPM}, where each time-step is trained independently. However, the lack of coordination among time-steps can lead to error accumulation and suboptimal reconstruction quality. To address this issue, we employ a two-stage training strategy. As shown in Fig. \ref{FSFT}(a), we first train each time-step $n$ independently, allowing the model to learn to remove noise and residuals at each step. The optimization objective consists of the rate-distortion loss, codebook loss \cite{l_codebook} and noise estimation loss:
\begin{equation}
\begin{split}
    \mathcal{L}_{stage\ I} = \underbrace{\Vert sg(\boldsymbol{l}_p) - \hat{\boldsymbol{l}}_p \Vert_2^2 + \beta \Vert sg(\hat{\boldsymbol{l}}_p) - \boldsymbol{l}_p \Vert_2^2}_{codebook\ loss \ \mathcal{L}_{cb}} + \\ \underbrace{\lambda_{r}\Vert \boldsymbol{z}_0 - \boldsymbol{z}_c \Vert_2^2 + R(\hat{\boldsymbol{y}})}_{rate-distortion\ loss \ \mathcal{L}_{rd}} + \lambda_{r} \mathcal{L}_{ne},
\end{split}
\end{equation}
where $\lambda_r$ is the hyper-parameter that controls the trade-off, $R(\cdot)$ denotes the estimated rate,  $sg(\cdot)$ denotes the stop-gradient operator, and $\beta=0.25$. Thanks to the proposed relay residual diffusion framework, we can achieve high-quality reconstruction in fewer than 5 denoising steps. This efficiency allows us to fine-tune the model using the entire reconstruction process with limited computational resources.

To this end, we further employ a fixed-step fine-tuning strategy to eliminate the discrepancy between the training and inference phases. As shown in Fig. \ref{FSFT}(b), in each training step, we utilize spaced DDPM sampling \cite{SpacedSampler} with $L$ fixed time-steps to reconstruct the noise-free latent features $\hat{\boldsymbol{z}}_0$ from the starting point $\boldsymbol{z_N}$ and map $\hat{\boldsymbol{z}}_0$ back to the pixel space $\hat{\boldsymbol{x}}=\mathcal{D}(\hat{\boldsymbol{z}}_0)$. The loss function used in this stage is as follows:
\begin{equation}
\begin{split}
    L_{stage\ II} = \lambda_{r} \Bigl(\Vert \boldsymbol{x} - \hat{\boldsymbol{x}} \Vert_2^2 + \lambda_{lpips} \mathcal{L}_{lpips}(\boldsymbol{x}, \hat{\boldsymbol{x}})\Bigr) + \\ \mathcal{L}_{rd} + \mathcal{L}_{cb} + \lambda_r \Vert \boldsymbol{z}_0 - \hat{\boldsymbol{z}}_0\Vert^2_2,
\end{split}
\label{loss_II}
\end{equation}
where $\mathcal{L}_{lpips}$ denotes the LPIPS loss and $\lambda_{lpips}=0.5$ is the weight of the LPIPS loss. By fine-tuning the model using the entire reconstruction process, we achieve significant performance improvement.


\subsection{Controllable Detail Generation}
Although the fixed-step fine-tuning strategy significantly improves reconstruction quality, it requires a fixed number of denoising steps in the inference phase, making it impossible to achieve a trade-off between smoothness and sharpness by adjusting the number of denoising steps \cite{DiffEIC}. To address this limitation, we introduce a controllable detail generation method that allows us to dynamically balance smoothness and sharpness without being constrained by the fixed-step requirement, which enables more versatile and user-specific image reconstructions. 

\begin{figure*}[!t]\scriptsize
\centering
\includegraphics[width=0.95\textwidth]{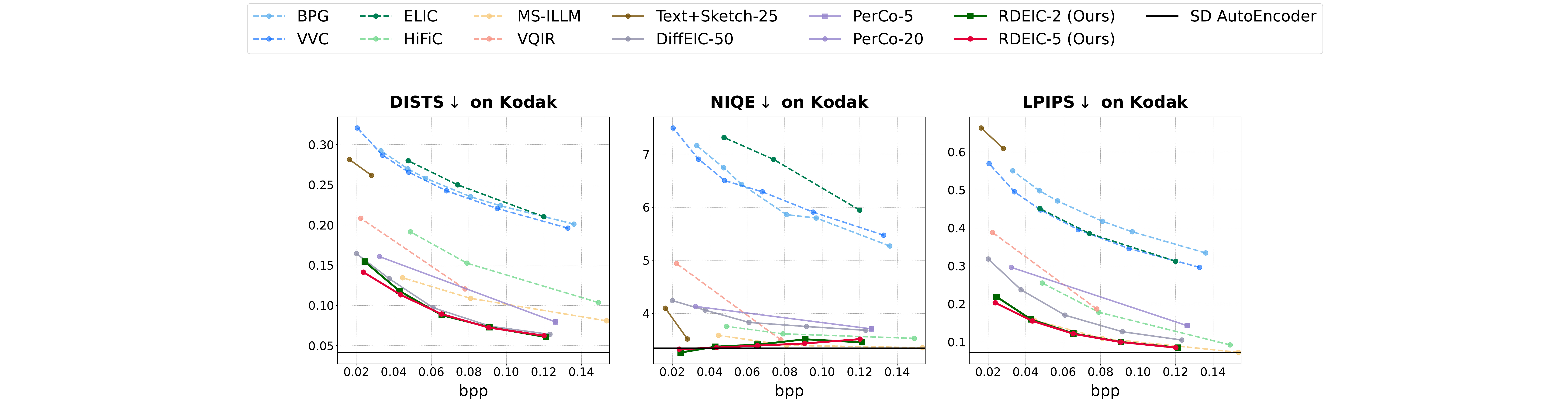}
\\ \vspace{0.1cm}
\includegraphics[width=0.95\textwidth]{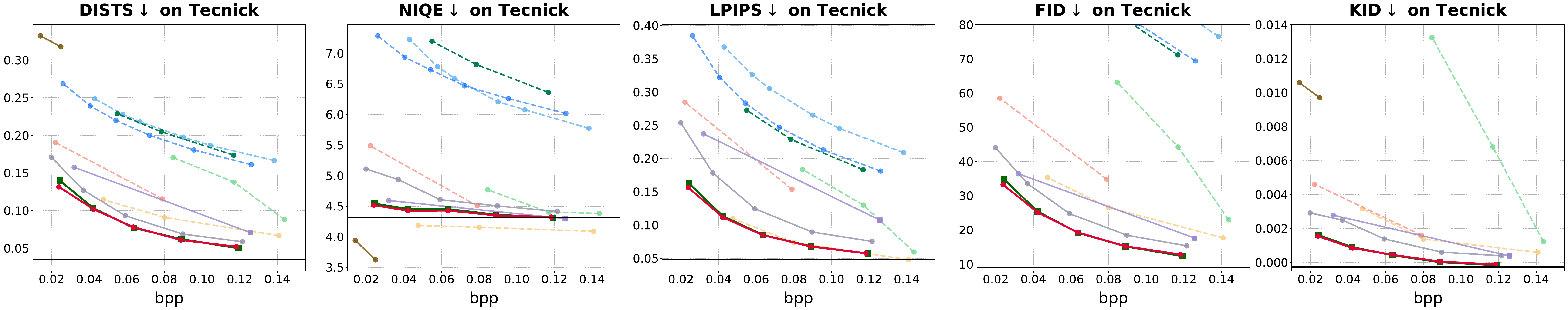}
\\ \vspace{0.1cm}
\includegraphics[width=0.95\textwidth]{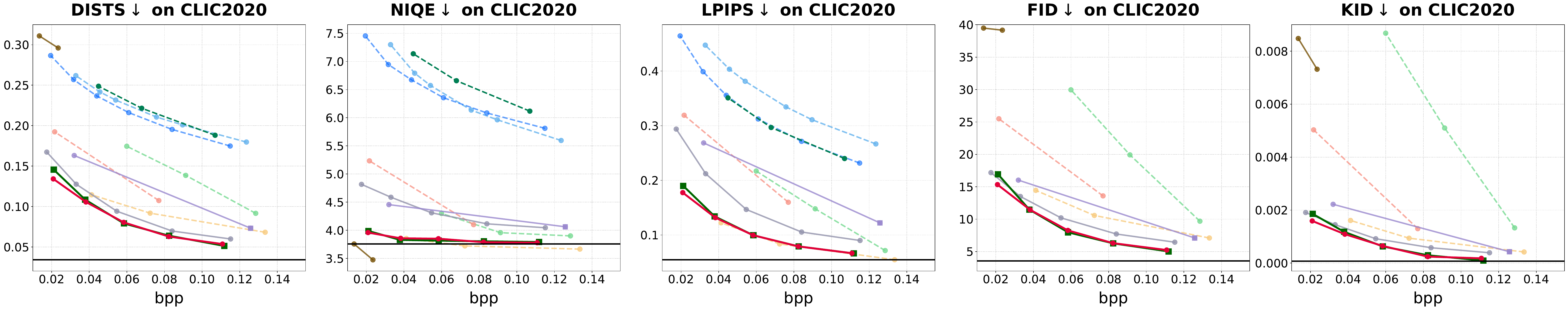}
\caption{{Quantitative comparisons with state-of-the-art methods in terms of perceptual metrics (DISTS $\downarrow$ / NIQE $\downarrow$ / LPIPS $\downarrow$ / FID $\downarrow$ / KID $\downarrow$ ) on the Kodak, Tecnick, and CLIC2020 datasets. Solid lines are used for diffusion-based methods, while dashed lines represent other methods.}}
\label{quant_results}
\end{figure*}

{In image generation tasks, Classifier-Free Guidance (CFG) \cite{cfg} is often used to enhance conditional generation by combining unconditional and conditional noise predictions:
\begin{equation}
    \hat{\epsilon} = \epsilon_{uncond} + \omega \cdot (\epsilon_{cond} - \epsilon_{uncond}),
\end{equation}
where $\epsilon_{cond}$ and $\epsilon_{uncond}$ denote the noise predictions with and without conditioning, respectively. $\omega$ is the guidance scale that controls the influence of the conditional input. 
As $\omega$ increases, the conditional signal is amplified, resulting in outputs that are sharper and more faithful to the conditional input.}
{Inspired by CFG, we adopt a similar principle in our reconstruction stage to control the trade-off between smoothness and sharpness. Specifically, we treat $\epsilon_{\text{sd}}(z_n, n)$ as the unconditional prediction from the pre-trained Stable Diffusion denoiser, and $\epsilon_{\theta}(z_n, c, n)$ as the conditional prediction guided by the conditional features $\boldsymbol{c}$. The final noise estimate is computed as:
\begin{equation}
    \hat{\epsilon}_n = \epsilon_{sd}(\boldsymbol{z}_n, n) + \lambda_s \cdot (\epsilon_\theta (\boldsymbol{z}_n, \boldsymbol{c}, n) - \epsilon_{sd}(\boldsymbol{z}_n, n)),
\end{equation}
where $\lambda_s$ plays the same role as the guidance scale $\omega$, allowing us to regulate the amount of high-frequency details introduced into the reconstructed image.} In the following experiments, we set $\lambda_s=1$ by default unless otherwise specified.
%
\section{Experiments}
\label{experiments}
\begin{figure*}[!t]\scriptsize
\centering
\includegraphics[width=0.95\textwidth]{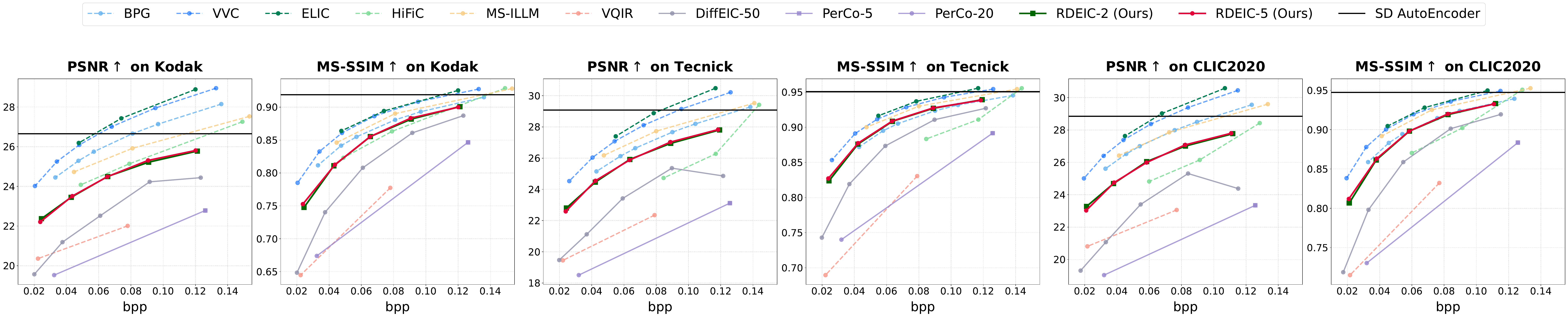}
\caption{{Quantitative comparisons with state-of-the-art methods in terms of fidelity (PSNR $\uparrow$ / MS-SSIM $\uparrow$) on the Kodak, Tecnick ,and CLIC2020 datasets. Since Text+Sketch \cite{Text+Sketch} has ignored the pixel-level fidelity of the reconstruction results, we do not report its rate-distortion performance.}}
\label{quant_distortion}
\end{figure*}

\subsection{Experimental Setup}
\subsubsection{Datasets} 
The proposed RDEIC is trained on the \textbf{LSDIR} \cite{LSDIR} dataset, which contains 84,911 high-quality images. For evaluation, we use three common benchmark datasets, i.e., the \textbf{Kodak} \cite{Kodak} dataset with 24 natural images of 768$\times$512 pixels, the \textbf{Tecnick} \cite{Tecnick} dataset with 140 images of 1200$\times$1200 pixels, and the \textbf{CLIC2020} \cite{CLIC2020} dataset with 428 high-quality images. For the Tecnick and CLIC2020 datasets, we resize the images so that the shorter dimension is equal to 768 and then center-crop them with 768$\times$768 spatial resolution \cite{CDC}. {The Kodak dataset was evaluated directly at its original resolution.}

\subsubsection{Implementation details}
We use Stable Diffusion 2.1-base\footnote{\url{https://huggingface.co/stabilityai/stable-diffusion-2-1-base}} as the specific implementation of stable diffusion. Throughout all our experiments, the weights of stable diffusion remain frozen. Similar to DiffEIC \cite{DiffEIC}, the control module in our RDEIC has the same encoder and middle block architecture as stable diffusion and reduces the channel number to 20\% of the original. The variance sequence $\{\beta_t\}_{t=1}^{T}$ used for adding noise is identical to that in Stable Diffusion. The number $N$ of denoising steps is set to 300. The size of codebook is set to 16,384. 

For training, we use the Adam \cite{Adam} optimizer with $\beta_1=0.9$ and $\beta_2=0.999$ for a total of 300K iterations. To achieve different compression ratios, we train five models with $\lambda_r$ selected from \{2, 1, 0.5, 0.25, 0.1\}. The batch size is set to 4. As described in Section \ref{FSFTS}, the training process is divided into two stages. \textit{1) Independent training.} During this stage, the initial learning rate is set to 1$\times 10^{-4}$ and images are randomly cropped to 512$\times$512 patches. We first train the proposed RDEIC with $\lambda_r = 2$ for 100K iterations. The learning rate is then reduced to 2$\times 10^{-5}$ and the model is trained with target $\lambda_r$ for another 100K iterations. \textit{2) Fixed-step fine-tuning}. In this stage, the learning rate is set to 2$\times 10^{-5}$. {To reduce the computational burden, images are randomly cropped to 256$\times$256 patches.} We fine-tune the model through the entire reconstruction process for 100K iterations. All experiments are conducted on a single NVIDIA GeForce RTX 4090 GPU.

\begin{figure*}[!t]\scriptsize
\begin{center}
\makebox[0.135\textwidth]{\textbf{Original}}
\makebox[0.135\textwidth]{\textbf{VVC}}
\makebox[0.135\textwidth]{\textbf{MS-ILLM}}
\makebox[0.135\textwidth]{\textbf{Text+Sketch}}
\makebox[0.135\textwidth]{\textbf{PerCo}}
\makebox[0.135\textwidth]{\textbf{DiffEIC}}
\makebox[0.135\textwidth]{\textbf{RDEIC (Ours)}}
\\ \vspace{0.1cm}
\includegraphics[width=0.135\textwidth]{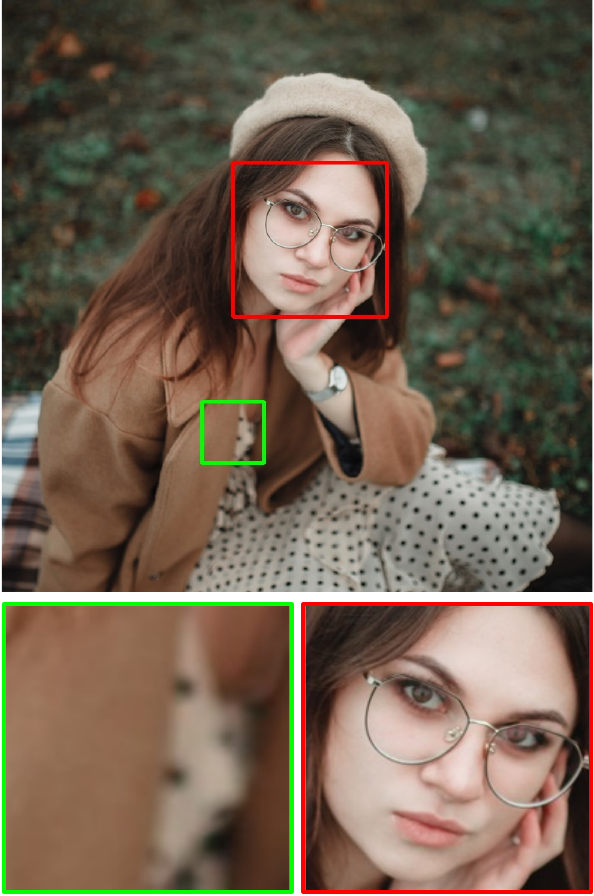}
\includegraphics[width=0.135\textwidth]{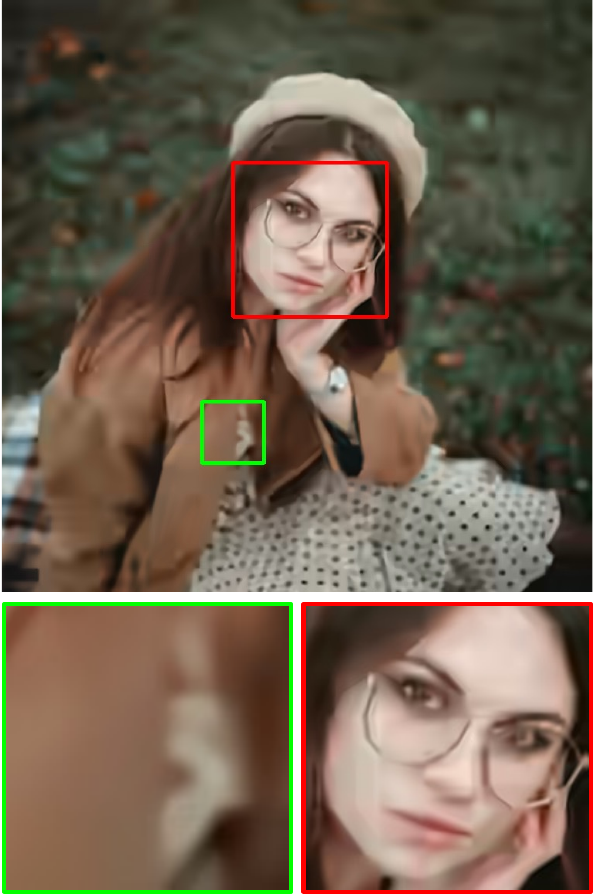}
\includegraphics[width=0.135\textwidth]{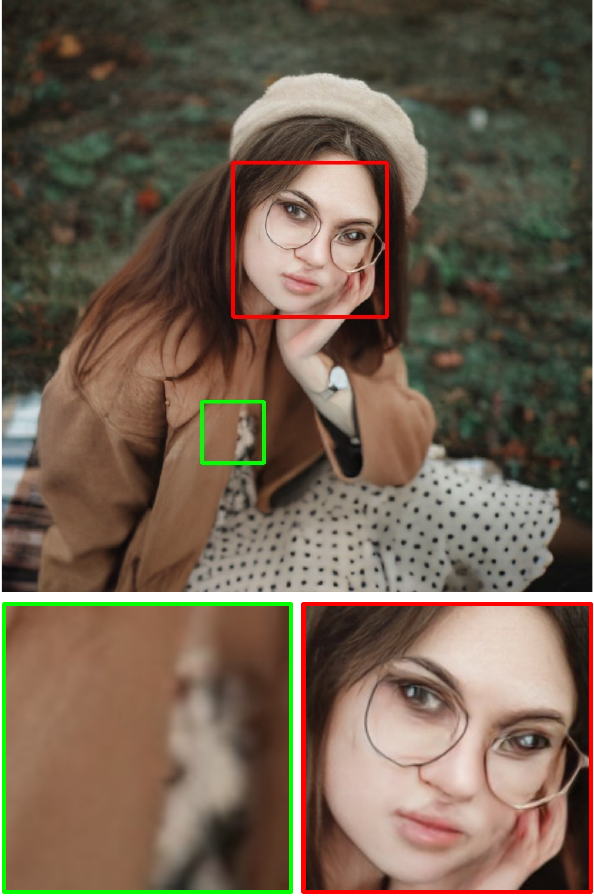}
\includegraphics[width=0.135\textwidth]{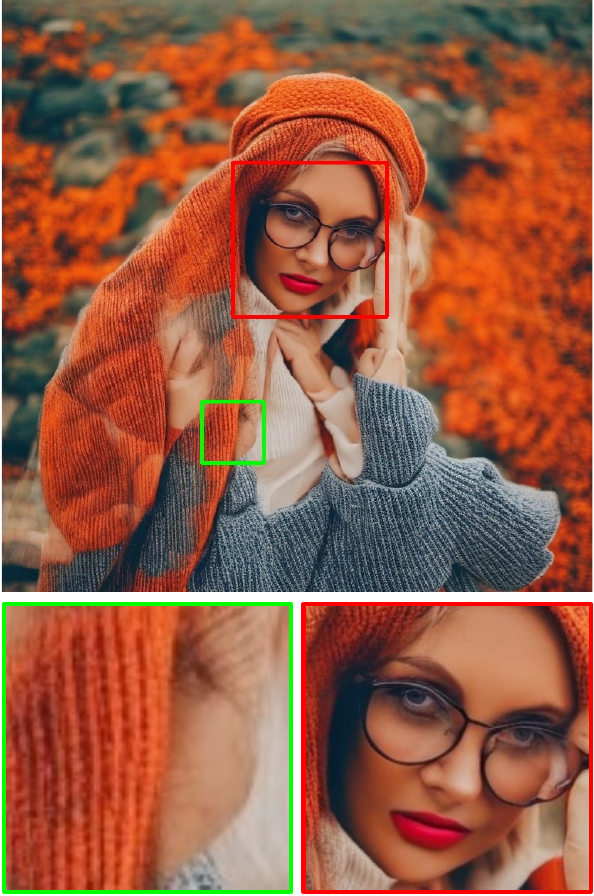}
\includegraphics[width=0.135\textwidth]{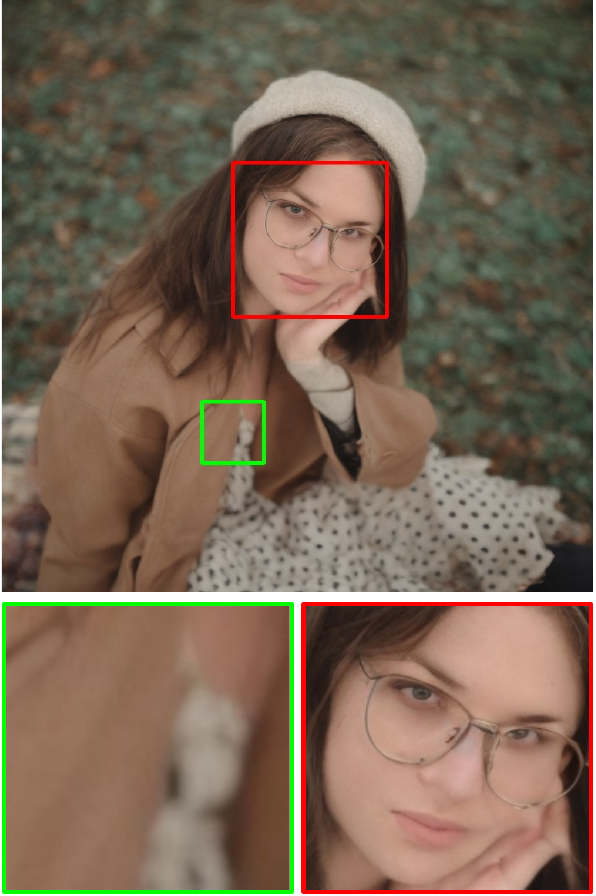}
\includegraphics[width=0.135\textwidth]{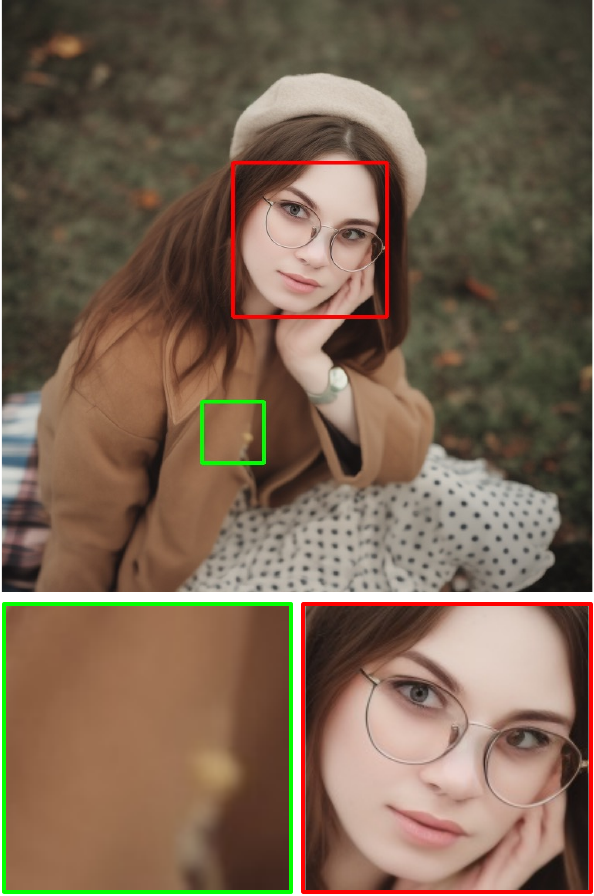}
\includegraphics[width=0.135\textwidth]{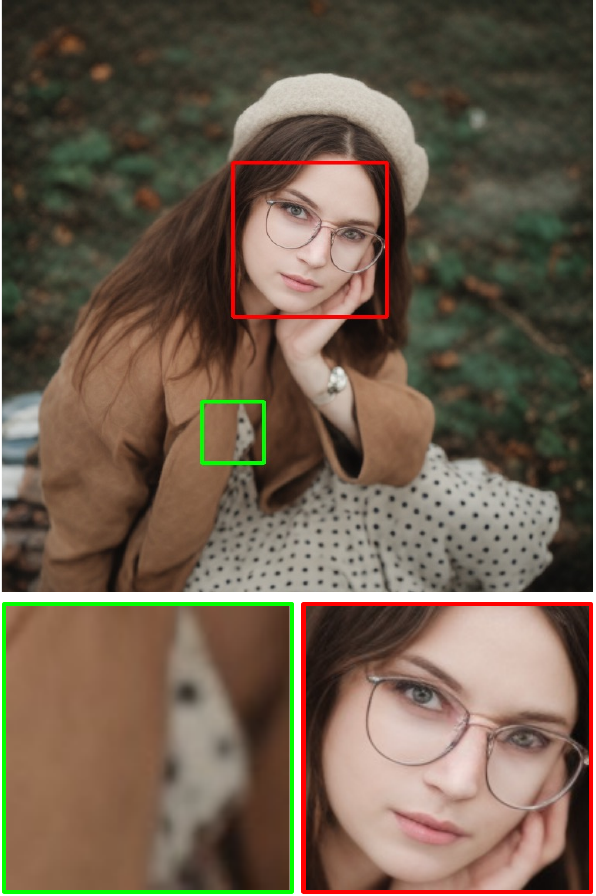}
\\ \vspace{0.1cm}
\makebox[0.135\textwidth]{\textbf{bpp / DISTS$\downarrow$}}
\makebox[0.135\textwidth]{\textbf{0.0339 / 0.2063}}
\makebox[0.135\textwidth]{\textbf{0.0385 / 0.0857}}
\makebox[0.135\textwidth]{\textbf{0.0182 / 0.2939}}
\makebox[0.135\textwidth]{\textbf{0.0320 / 0.1313}}
\makebox[0.135\textwidth]{\textbf{0.0236 / 0.1265}}
\makebox[0.135\textwidth]{\textbf{0.0192 / 0.1014}}
\\ \vspace{0.1cm}
\includegraphics[width=0.135\textwidth]{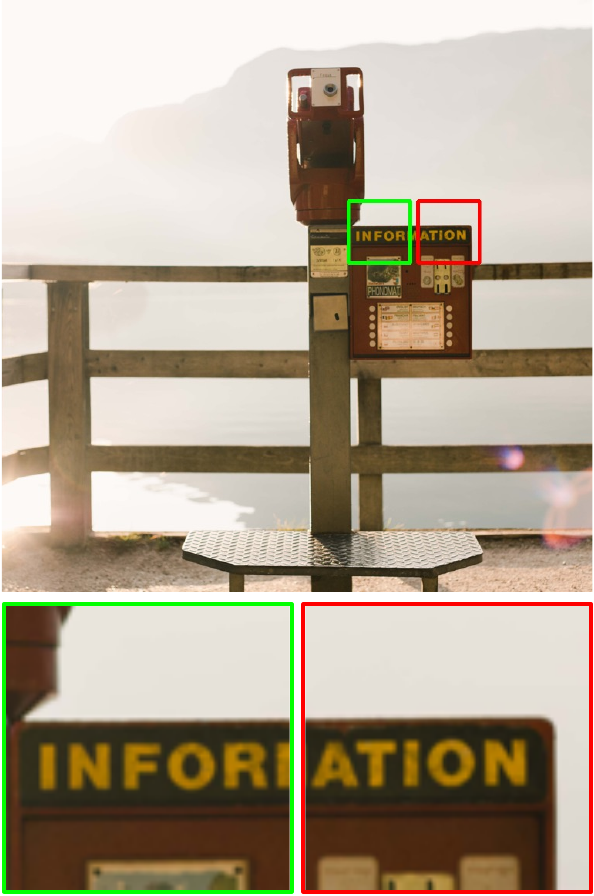}
\includegraphics[width=0.135\textwidth]{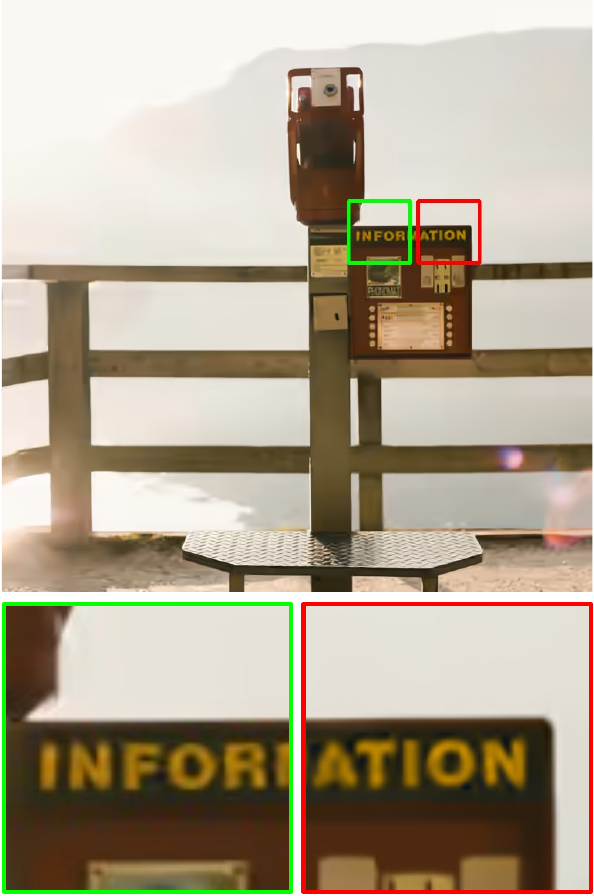}
\includegraphics[width=0.135\textwidth]{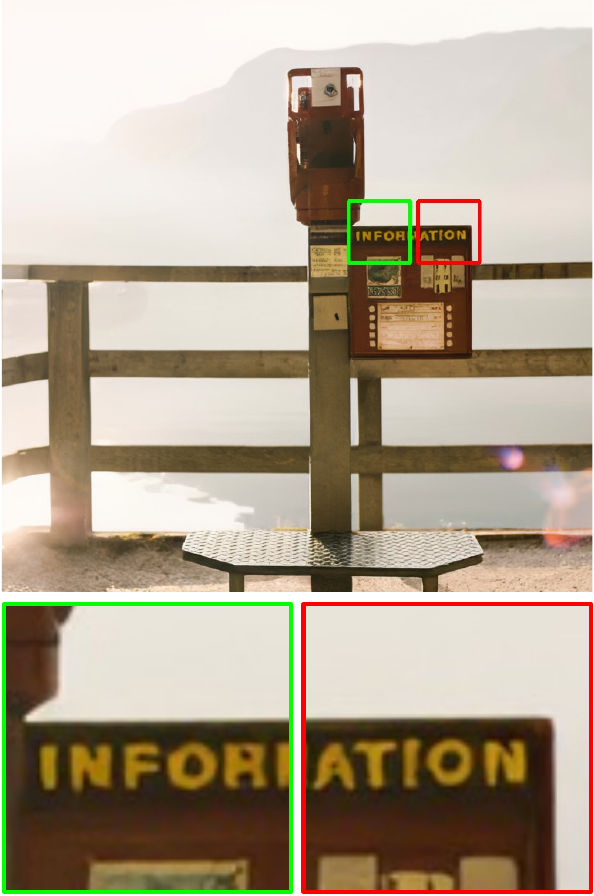}
\includegraphics[width=0.135\textwidth]{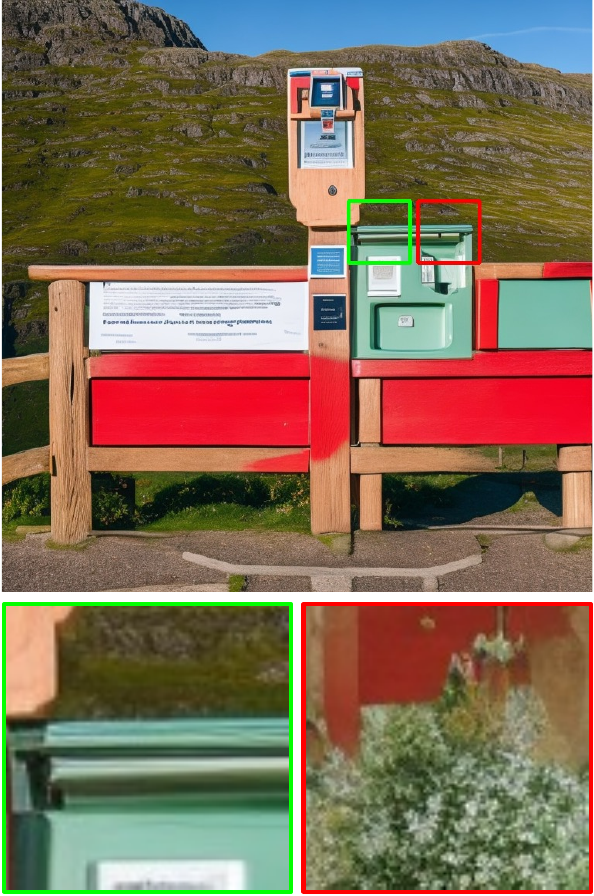}
\includegraphics[width=0.135\textwidth]{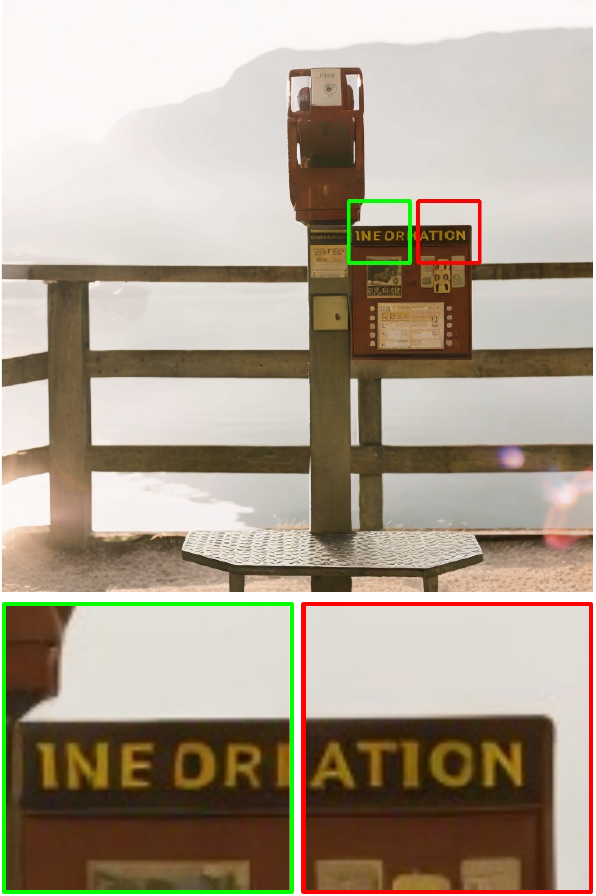}
\includegraphics[width=0.135\textwidth]{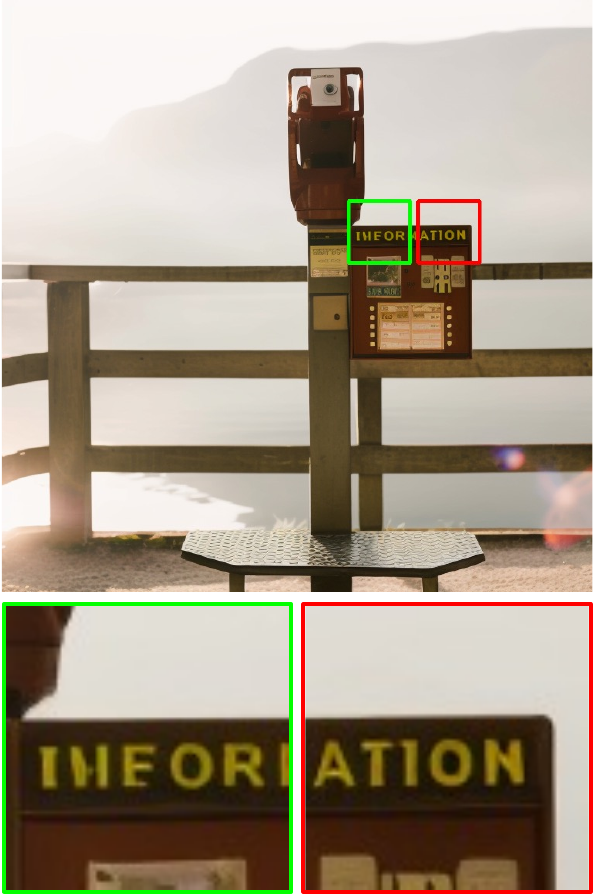}
\includegraphics[width=0.135\textwidth]{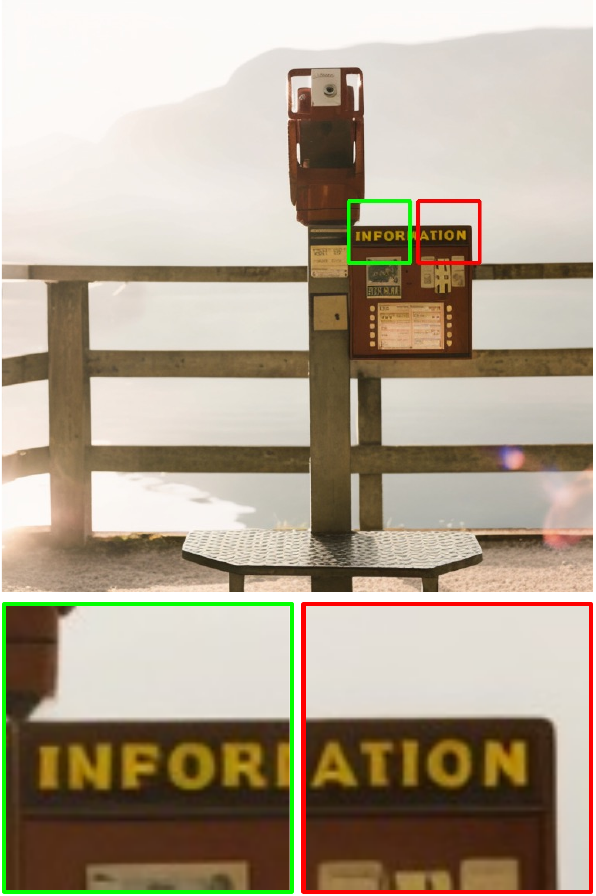}
\\ \vspace{0.1cm}
\makebox[0.135\textwidth]{\textbf{bpp / DISTS$\downarrow$}}
\makebox[0.135\textwidth]{\textbf{0.0663 / 0.1238}}
\makebox[0.135\textwidth]{\textbf{0.0783 / 0.0569}}
\makebox[0.135\textwidth]{\textbf{0.0165 / 0.3409}}
\makebox[0.135\textwidth]{\textbf{0.1256 / 0.0567}}
\makebox[0.135\textwidth]{\textbf{0.0605 / 0.0624}}
\makebox[0.135\textwidth]{\textbf{0.0591 / 0.0485}}
\end{center}
\caption{Visual comparisons of our method to baselines on the CLIC2020 dataset. Compared to other methods, our method produces more realistic and faithful reconstructions. {For our RDEIC, the first example corresponds to the result of RDEIC-5, while the second example corresponds to RDEIC-2.}}
\label{qualitative comparisons}
\end{figure*}

\subsubsection{Metrics} For quantitative evaluation, we employ several established metrics to measure the visual quality of the reconstructed images, including reference perceptual metrics \textbf{LPIPS} \cite{LPIPS}, \textbf{DISTS} \cite{DISTS}, \textbf{FID} \cite{FID} and \textbf{KID} \cite{KID} and no-reference perceptual metric \textbf{NIQE} \cite{NIQE}. We also employ distortion metrics \textbf{PSNR} and \textbf{MS-SSIM} \cite{MS-SSIM} to measure the fidelity of reconstructions. Note that FID and KID are calculated on 256$\times$256 patches according to \cite{HiFiC}. Since the Kodak dataset is too small to reliably calculate FID and KID scores, we do not report these results for this dataset.

\subsection{Comparisons With State-of-the-art Methods}
We compare the proposed RDEIC with several representative extreme image compression methods, including the traditional standards: BPG \cite{BPG} and VVC \cite{VVC}; VAE-based method: ELIC \cite{ELIC}; GANs-based methods: HiFiC \cite{HiFiC}, MS-ILLM \cite{MS-ILLM}, and VQIR \cite{VQIR}; and diffusion-based methods: Text+Sketch \cite{Text+Sketch}, PerCo \cite{PerCo},and DiffEIC \cite{DiffEIC}. {For our RDEIC, we present the performance curves of models fine-tuned with 2 and 5 fixed denoising steps, i.e., RDEIC-2 and RDEIC-5.}

\begin{table*}[htbp]
\renewcommand{\arraystretch}{1.1}
\caption{Encoding and decoding time (in seconds) on Kodak dataset. Decoding time is divided into the time spent in the denoising stage and the time spent in the remaining parts. {BD-Rate is evaluated on the CLIC2020 dataset using DISTS as the metric, with VVC \cite{VVC} serving as the anchor.
The best and second best results are highlighted in \textbf{bold} and \underline{underline}.} The testing platform is RTX4090.}
\label{comparition_time}
\centering
\begin{tabular}{llc|rrrc} 
\toprule
\multirow{2}{*}{\textbf{Types}} &\multirow{2}{*}{\textbf{Methods}} &\multirow{2}{*}{\textbf{Denoising Step}}  &\multirow{2}{*}{\textbf{Encoding Time}} &\multicolumn{2}{c}{\textbf{Decoding time}} &\multirow{2}{*}{{\textbf{BD-Rate(\%)$\downarrow$}}}\\ \cline{5-6}
& & & &{\textbf{Denoising Time}} &{\textbf{Remaining Time}} \\ \midrule
{VAE-based} & {ELIC}  & {--}& {0.056 $\pm$ 0.006} & \multicolumn{1}{c}{{\ \ --}} & {0.081 $\pm$ 0.011} &{20.06}\\ \midrule
\multirow{3}{*}{GAN-based} &HiFiC &-- & 0.038 $\pm$ 0.004 &\multicolumn{1}{c}{{\ \ --}} & {0.059 $\pm$ 0.004} &{-48.11}\\
&MS-ILLM &-- & 0.038 $\pm$ 0.004 &\multicolumn{1}{c}{{\ \ --}} & {0.059 $\pm$ 0.004} &{-88.72}\\
&VQIR &-- & 0.050 $\pm$ 0.003 & \multicolumn{1}{c}{{\ \ --}} & {0.179 $\pm$ 0.005} &{-75.49}\\ \midrule
\multirow{6}{*}{Diffusion-based} &Text+Sketch &25 & 62.045 $\pm$ 0.516  &{8.483 $\pm$ 0.344} & {4.030 $\pm$ 0.469} &{111.94}\\
&DiffEIC &50 & 0.128 $\pm$ 0.005 & {4.342 $\pm$ 0.013} & {0.228 $\pm$ 0.026} &{-86.86}\\ \cline{2-7}
&\multirow{2}{*}{PerCo} &5 &0.236 $\pm$ 0.040 & {0.623 $\pm$ 0.003} &{0.186 $\pm$ 0.002} &{-76.70}\\
& &20 &0.236 $\pm$ 0.040 & {2.495 $\pm$ 0.009} &{0.186 $\pm$ 0.002} &{-76.70}\\ \cline{2-7}
&\multirow{2}{*}{RDEIC (Ours)} &2 &0.119 $\pm$ 0.003 & {0.173 $\pm$ 0.001} &{0.198 $\pm$ 0.003} &{\underline{-89.44}}\\
& &5 &0.119 $\pm$ 0.003 & {0.434 $\pm$ 0.002} &{0.198 $\pm$ 0.003} &{\textbf{-94.27}}\\
\bottomrule
\end{tabular}
\end{table*}

\begin{table*}[!t]
\caption{The impact of Relay Residual Diffusion (RRD) and Fixed-Step Fine-Tune (FSFT) on performance. BD-Rate is evaluated on CLIC2020 dataset, using DiffEIC \cite{DiffEIC} as the anchor. DS denotes the number of denoising steps.}
\label{ablations_bd}
\centering
\begin{tabular}{lcrrrrcr}
\toprule
\multirow{2}{*}{\textbf{Methods}} &\multirow{2}{*}{\textbf{DS}} &\multicolumn{4}{c}{\textbf{BD-Rate (\%) $\downarrow$}} &\multirow{2}{*}{\textbf{Denoising Time}} &\multirow{2}{*}{\textbf{Speedup}}\\ \cline{3-6}
& &DISTS &LPIPS &PSNR &MS-SSIM\\ \midrule
DiffEIC & 50 & 0 &0 &0 &0 &4.342 $\pm$ 0.013 &1$\times$\\
Baseline & 50 &1.57 &4.88 &9.72 &3.58 & 4.349 $\pm$ 0.013 & 1$\times$\\ \midrule
\multirow{2}{*}{+RRD} & 5 & 12.43 & -4.50 & -47.45 & -18.70 &0.434 $\pm$ 0.002 & 10$\times$ \\
 & 2 & 36.33 & 13.83 & -56.75 & -26.21 & 0.173 $\pm$ 0.001 & 25$\times$\\ \midrule
\multirow{2}{*}{RDEIC (+RRD+FSFT)} & 5 &-18.02 &-40.79 &-61.11 &-32.73 &0.434 $\pm$ 0.002 & 10$\times$\\
 & 2 &-15.08 &-39.07 &-61.27 &-31.89 & 0.173 $\pm$ 0.001 & 25$\times$\\
\bottomrule
\end{tabular}
\end{table*}

\subsubsection{Quantitative Comparisons}
\label{Quantitative}
Fig. \ref{quant_results} shows the rate-perception curves of the proposed and compared methods across the three datasets. It can be observed that the proposed RDEIC demonstrates superior performance across different perceptual metrics compared to other methods, particularly achieving optimal results in DISTS, FID, and KID. For the distortion metrics, as shown in Fig. \ref{quant_distortion}, RDEIC significantly outperforms other diffusion-based methods, underscoring its superiority in maintaining consistency. {RDEIC-2 and RDEIC-5 achieve comparable performance across most metrics. However, at the lowest bitrate setting (i.e., $\lambda_r = 0.1$), using more denoising steps results in better perceptual quality, as reflected in the perceptual metrics.}
Moreover, we report the performance of the SD autoencoder in Fig. \ref{quant_results} and Fig. \ref{quant_distortion} (see the black horizontal line), which represents the upper bound of RDEIC's performance, {as it reflects the best achievable reconstruction quality without compression}. Compared to DiffEIC \cite{DiffEIC}, which is also based on stable diffusion, RDEIC is significantly closer to this performance upper limit.

\subsubsection{Qualitative Comparisons}
Fig. \ref{exhibition} and Fig. \ref{qualitative comparisons} provides visual comparisons among the evaluated methods at extremely low bitrates. VVC \cite{VVC} and MS-ILLM \cite{MS-ILLM} excel at reconstructing structural information, such as text, but falls significantly short in preserving textures and fine details. Diffusion-based Text+Sketch \cite{Text+Sketch}, PerCo \cite{PerCo} and DiffEIC \cite{DiffEIC} achieve realistic reconstruction at extremely low bitrates but often generate details and structures that are inconsistent with the original image. In comparison, the proposed RDEIC produces reconstructions with higher visual quality, fewer artifacts, and more faithful details.

\subsubsection{Complexity Comparisons}
Table \ref{comparition_time} summarizes the average encoding/decoding times along with standard deviations for different methods on the Kodak dataset. For diffusion-based methods, decoding time is divided into denoising time and remaining time. Due to relay on stable diffusion, diffusion-based extreme image compression methods have higher encoding and decoding complexity than other learned-based methods. By reducing the number of denoising steps required for reconstruction, the denoising time of RDEIC is significantly lower than that of other diffusion-based methods. For instance, compared to DiffEIC \cite{DiffEIC}, our RDEIC is approximately 10$\times$ to 25$\times$ faster in terms of denoising time.
\section{Analysis and Discussions}
\label{analysis}
To provide a more comprehensive analysis of the proposed method, we conduct ablation studies. For the baseline, we employ the same diffusion framework as DiffEIC \cite{DiffEIC}, where the denoising process starts from pure noise. As shown in Table \ref{ablations_bd}, our baseline performs similarly to DiffEIC \cite{DiffEIC}, indicating that {the performance improvements of our method primarily stem from the proposed relay residual diffusion (RRD) framework and fixed-step fine-tuning (FSFT) strategy}, while the choice of compression module has minimal impact on the overall performance.

\begin{figure}[t]\scriptsize
\centering
\makebox[0.118\textwidth]{\textbf{(a) Original}}
\makebox[0.118\textwidth]{\textbf{(b) Baseline-2}}
\makebox[0.118\textwidth]{\textbf{(c) +RRD-2}}
\makebox[0.118\textwidth]{\textbf{(d) +RRD+FSFT-2}}
\\ \vspace{0.1cm}
\includegraphics[width=0.118\textwidth]{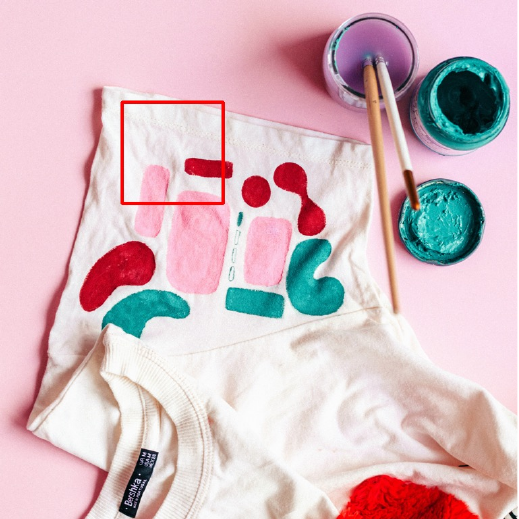}
\includegraphics[width=0.118\textwidth]{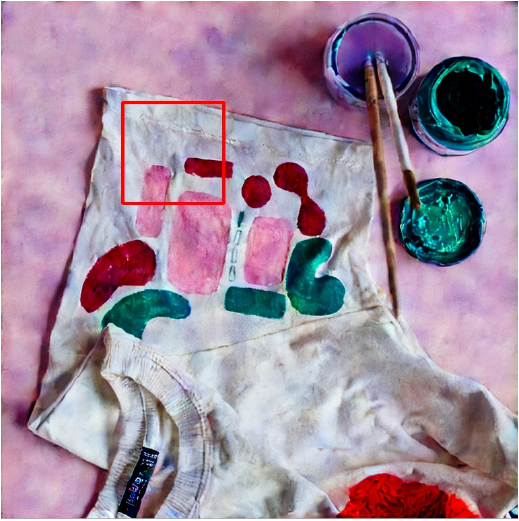}
\includegraphics[width=0.118\textwidth]{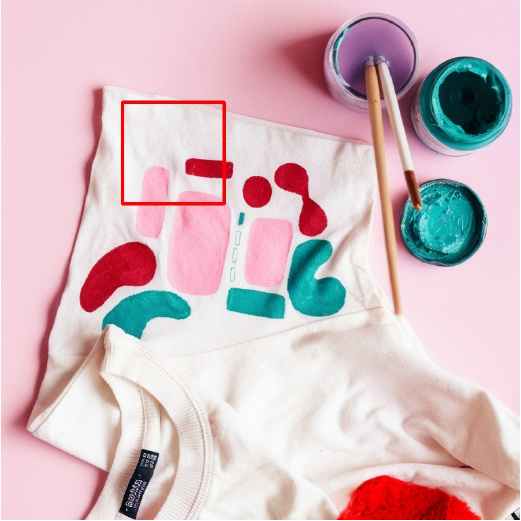}
\includegraphics[width=0.118\textwidth]{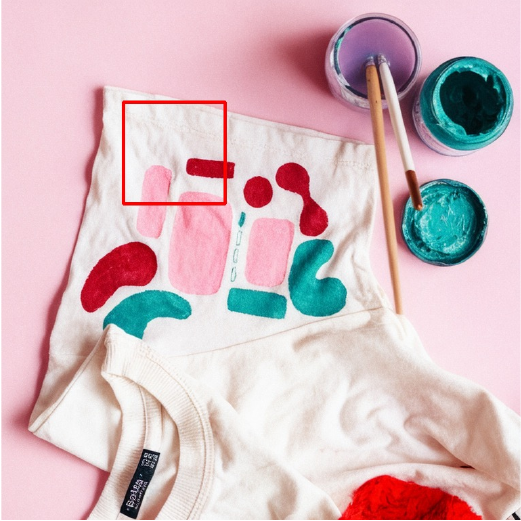}
\\ \vspace{0.05cm}
\includegraphics[width=0.118\textwidth]{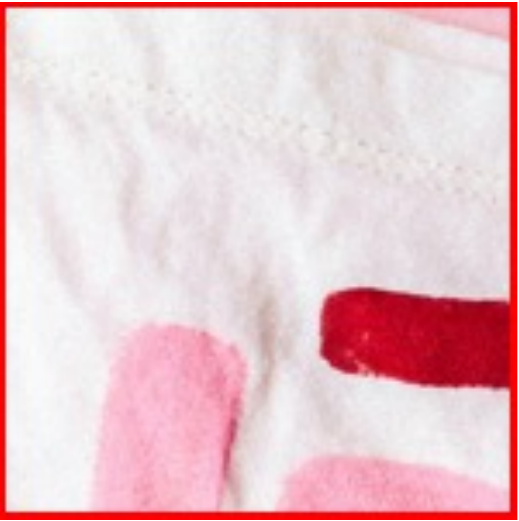}
\includegraphics[width=0.118\textwidth]{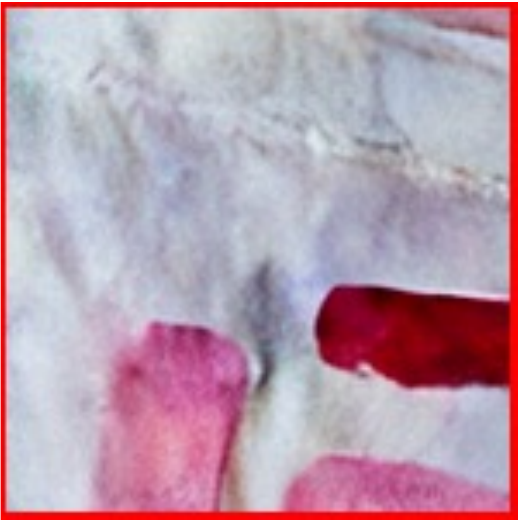}
\includegraphics[width=0.118\textwidth]{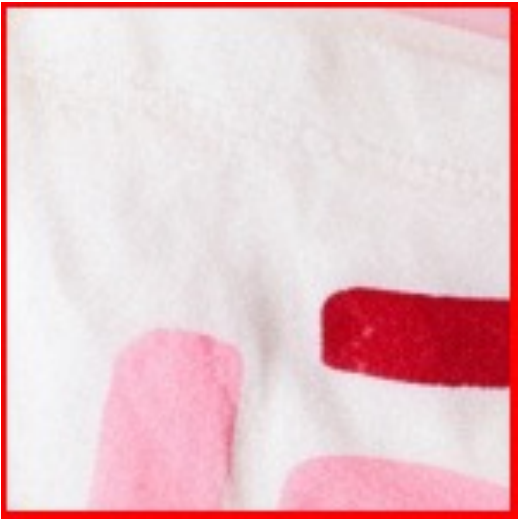}
\includegraphics[width=0.118\textwidth]{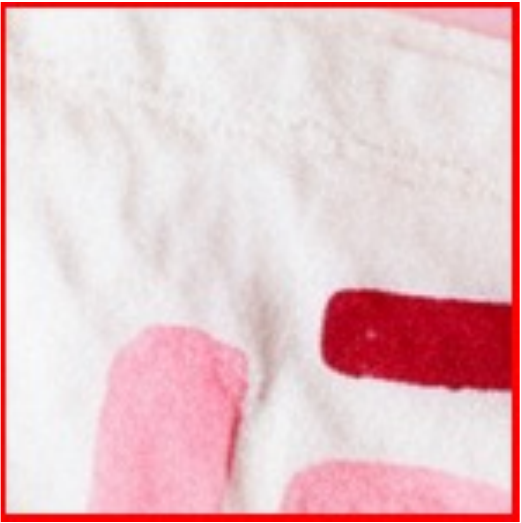}
\\ \vspace{0.1cm}
\makebox[0.118\textwidth]{\textbf{bpp / DISTS$\downarrow$}}
\makebox[0.118\textwidth]{\textbf{0.0996 / 0.1682}}
\makebox[0.118\textwidth]{\textbf{0.0815 / 0.0506}}
\makebox[0.118\textwidth]{\textbf{0.0767 / 0.0426}}
\caption{{Visual comparisons of each component in the proposed method.}}
\label{ablations_1}
\end{figure}

\begin{figure*}[!t]\scriptsize
\centering
\makebox[0.132\textwidth]{\textbf{Original}}
\makebox[0.132\textwidth]{\textbf{$\lambda_s$=0.0}}
\makebox[0.132\textwidth]{\textbf{$\lambda_s$=0.6}}
\makebox[0.132\textwidth]{\textbf{$\lambda_s$=0.8}}
\makebox[0.132\textwidth]{\textbf{$\lambda_s$=1.0}}
\makebox[0.132\textwidth]{\textbf{$\lambda_s$=1.3}}
\makebox[0.132\textwidth]{\textbf{$\lambda_s$=1.5}}
\\ \vspace{0.1cm}
\includegraphics[width=0.132\textwidth]{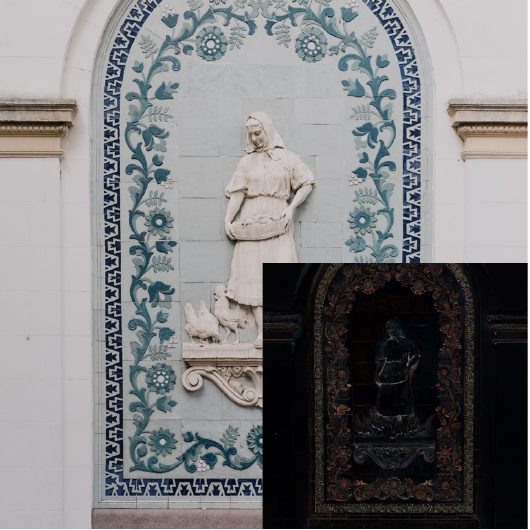}
\includegraphics[width=0.132\textwidth]{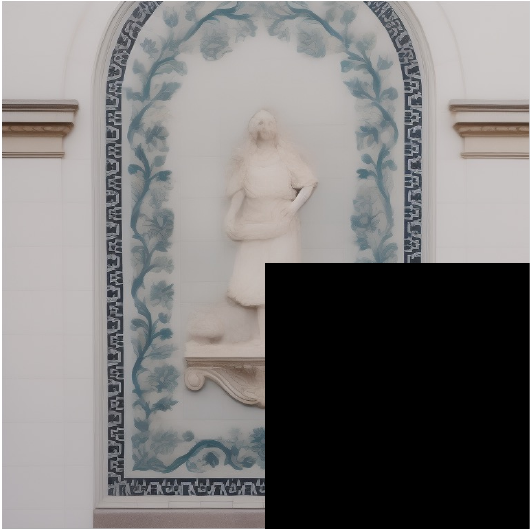}
\includegraphics[width=0.132\textwidth]{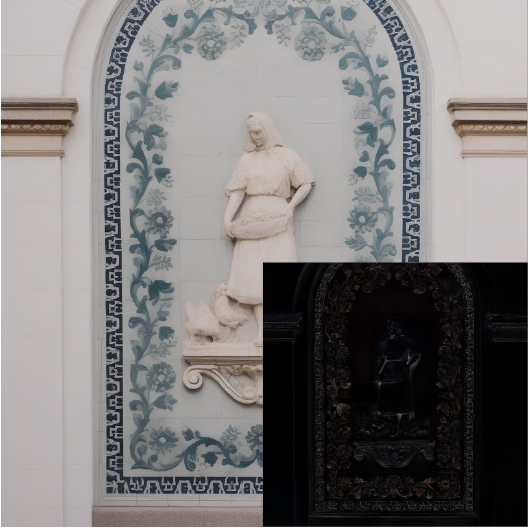}
\includegraphics[width=0.132\textwidth]{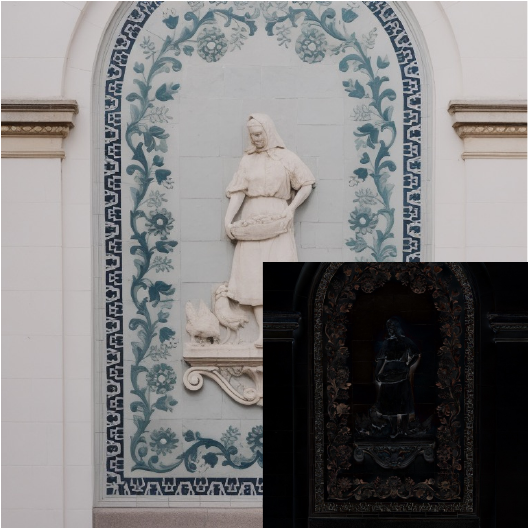}
\includegraphics[width=0.132\textwidth]{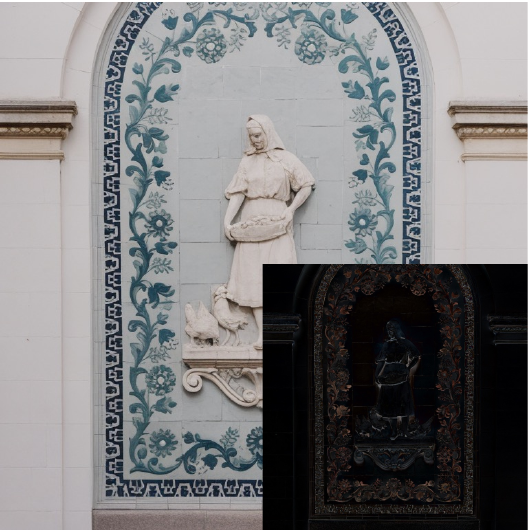}
\includegraphics[width=0.132\textwidth]{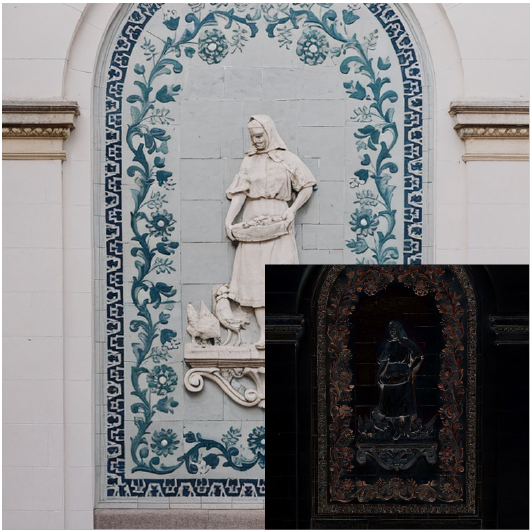}
\includegraphics[width=0.132\textwidth]{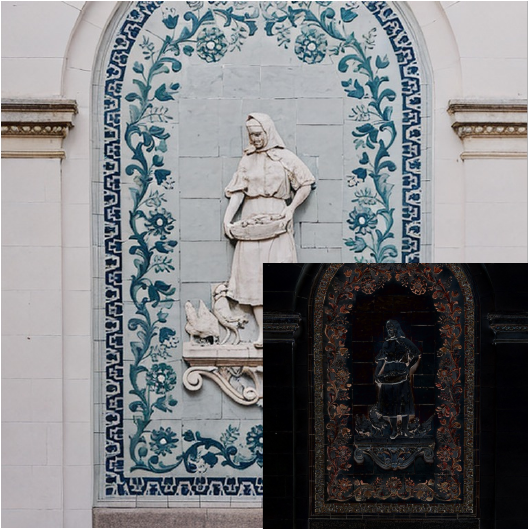}
\\
\makebox[0.132\textwidth]{\textbf{DISTS$\downarrow$ / PSNR$\uparrow$}}
\makebox[0.132\textwidth]{\textbf{0.2279 / 21.01}}
\makebox[0.132\textwidth]{\textbf{0.1223 / 23.47}}
\makebox[0.132\textwidth]{\textbf{0.0808 / 24.25}}
\makebox[0.132\textwidth]{\textbf{0.0687 / 24.39}}
\makebox[0.132\textwidth]{\textbf{0.0979 / 23.21}}
\makebox[0.132\textwidth]{\textbf{0.1352 / 21.76}}
\\ \vspace{0.1cm}
\includegraphics[width=0.132\textwidth]{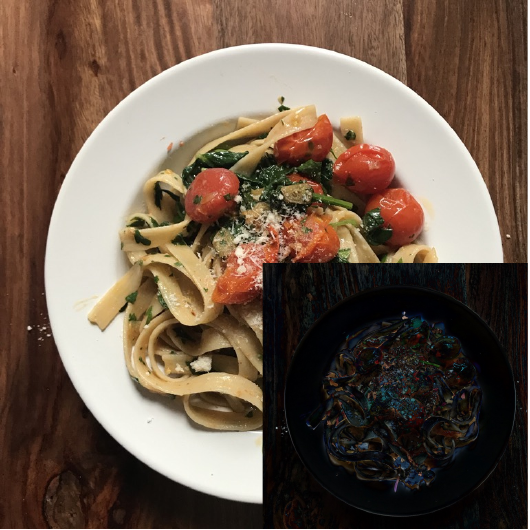}
\includegraphics[width=0.132\textwidth]{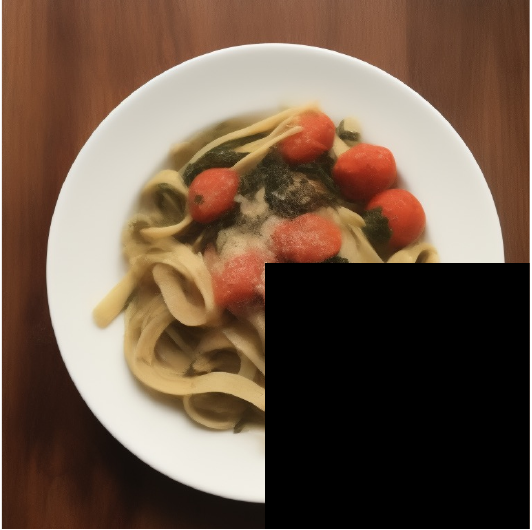}
\includegraphics[width=0.132\textwidth]{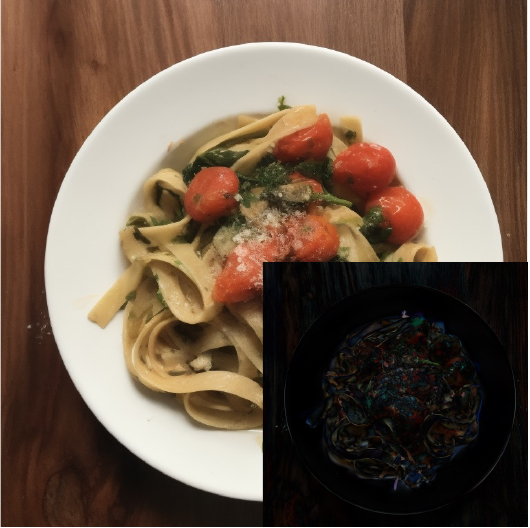}
\includegraphics[width=0.132\textwidth]{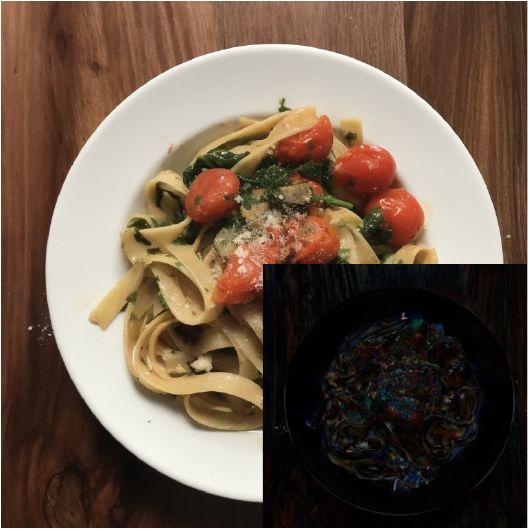}
\includegraphics[width=0.132\textwidth]{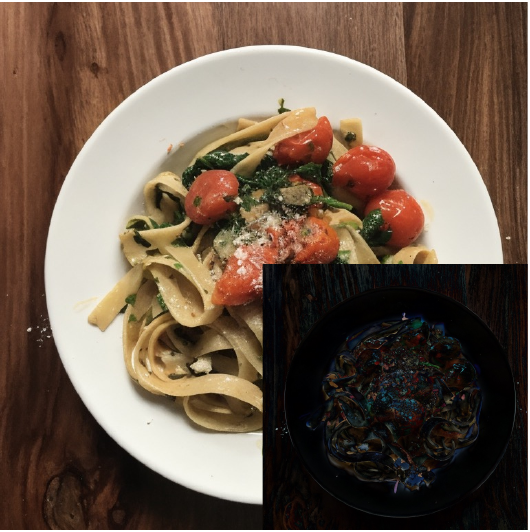}
\includegraphics[width=0.132\textwidth]{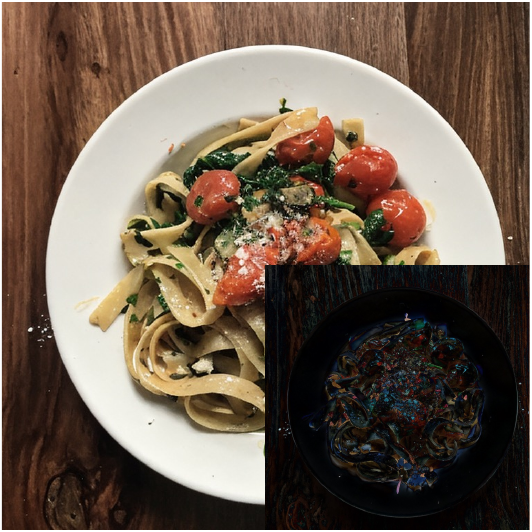}
\includegraphics[width=0.132\textwidth]{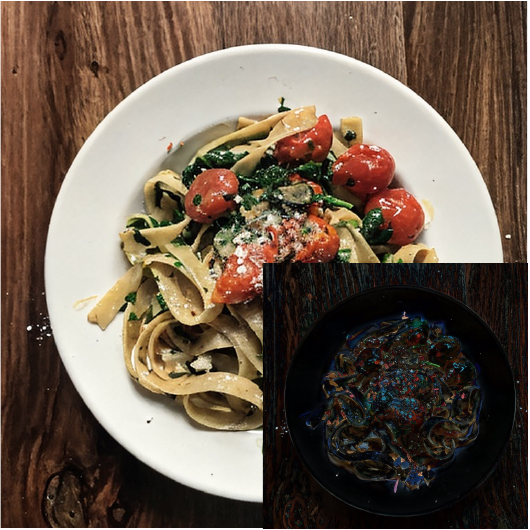}
\\
\makebox[0.132\textwidth]{\textbf{DISTS$\downarrow$ / PSNR$\uparrow$}}
\makebox[0.132\textwidth]{\textbf{0.2656 / 21.92}}
\makebox[0.132\textwidth]{\textbf{0.1345 / 25.36}}
\makebox[0.132\textwidth]{\textbf{0.0766 / 26.46}}
\makebox[0.132\textwidth]{\textbf{0.0555 / 26.64}}
\makebox[0.132\textwidth]{\textbf{0.0760 / 25.18}}
\makebox[0.132\textwidth]{\textbf{0.1168 / 23.43}}
\caption{{Balancing smoothness versus sharpness. These results are produced by RDEIC-2 trained with $\lambda_r = 1$. The bottom-right corner of each image displays the absolute difference between the reconstructed image and the reference image $\boldsymbol{x}_{base}$.}}
\label{diversity}
\end{figure*}

\subsection{Effectiveness of Relay Residual Diffusion}
We first investigate the effectiveness of our proposed relay residual diffusion framework. As shown in Table \ref{ablations_bd}, by incorporating the proposed relay residual diffusion framework, the method (+RRD) achieves better distortion metrics (PSNR and MS-SSIM) with only 2 or 5 denoising steps compared to the Baseline, which uses 50 denoising steps. The reason behind this is that starting from the compressed latent feature, instead of pure noise, avoids the error accumulation in the initial stage of the denoising process and provides a solid foundation for subsequent detail generation. {The slight drop in perceptual metrics (DISTS and LPIPS) is acceptable, as the relay residual diffusion framework requires significantly fewer denoising steps. Moreover, as shown in Fig. \ref{ablations_1}(b), when using only 2 denoising steps, the baseline method—which starts from pure noise—fails to completely remove the noise, resulting in noticeable artifacts in the reconstructed image. In contrast, our relay residual diffusion framework constructs the starting point from compressed latent features, enabling high-quality reconstruction even with 2 denoising steps, as shown in Fig. \ref{ablations_1}(c).} 

It is worth noting that, since the time required for the denoising stage is directly proportional to the number of denoising steps, incorporating the RRD framework reduces the denoising time by a factor of 10$\times$ to 25$\times$ compared to the baseline, as shown in Table \ref{ablations_bd}.

\subsection{Effectiveness of Fixed-step Fine-tuning}\
We further investigate the effectiveness of the proposed FSFT strategy. 
{Compared to existing diffusion-based extreme image compression methods, our proposed FSFT strategy fine-tunes the model using the entire reconstruction process and directly applies constraints on the final estimated noise-free latent features $\hat{\boldsymbol{z}}_0$ and the reconstructed image $\hat{\boldsymbol{x}}$, as shown in Eq. \ref{loss_II}. This strategy is aligned with the actual inference process and offers two primary advantages. On the one hand, by training along a reconstruction trajectory consistent with inference, FSFT effectively mitigates the mismatch between the training and inference phases, thereby enhancing the model’s stability and performance in practical scenarios. On the other hand, by directly constraining the latent features and reconstructed image, rather than relying on noise estimation errors, FSFT enables the model to explicitly learn the degradation patterns inherent in compressed latent features, facilitating more accurate recovery of lost image details.}

{As demonstrated in Table \ref{ablations_bd}, incorporating the FSFT strategy significantly improves performance across all evaluated metrics, confirming that FSFT effectively addresses the training-inference mismatch present in existing diffusion-based compression methods. Furthermore, as shown in Fig. \ref{ablations_1}(d), the model fine-tuned with the FSFT strategy demonstrates an enhanced ability to recover fine-grained image details, such as clothing folds.} It is worth noting that FSFT is a fine-tuning strategy that introduces no additional computational overhead during inference.

\begin{figure}[!t]\scriptsize
\centering
\includegraphics[width=0.45\textwidth]{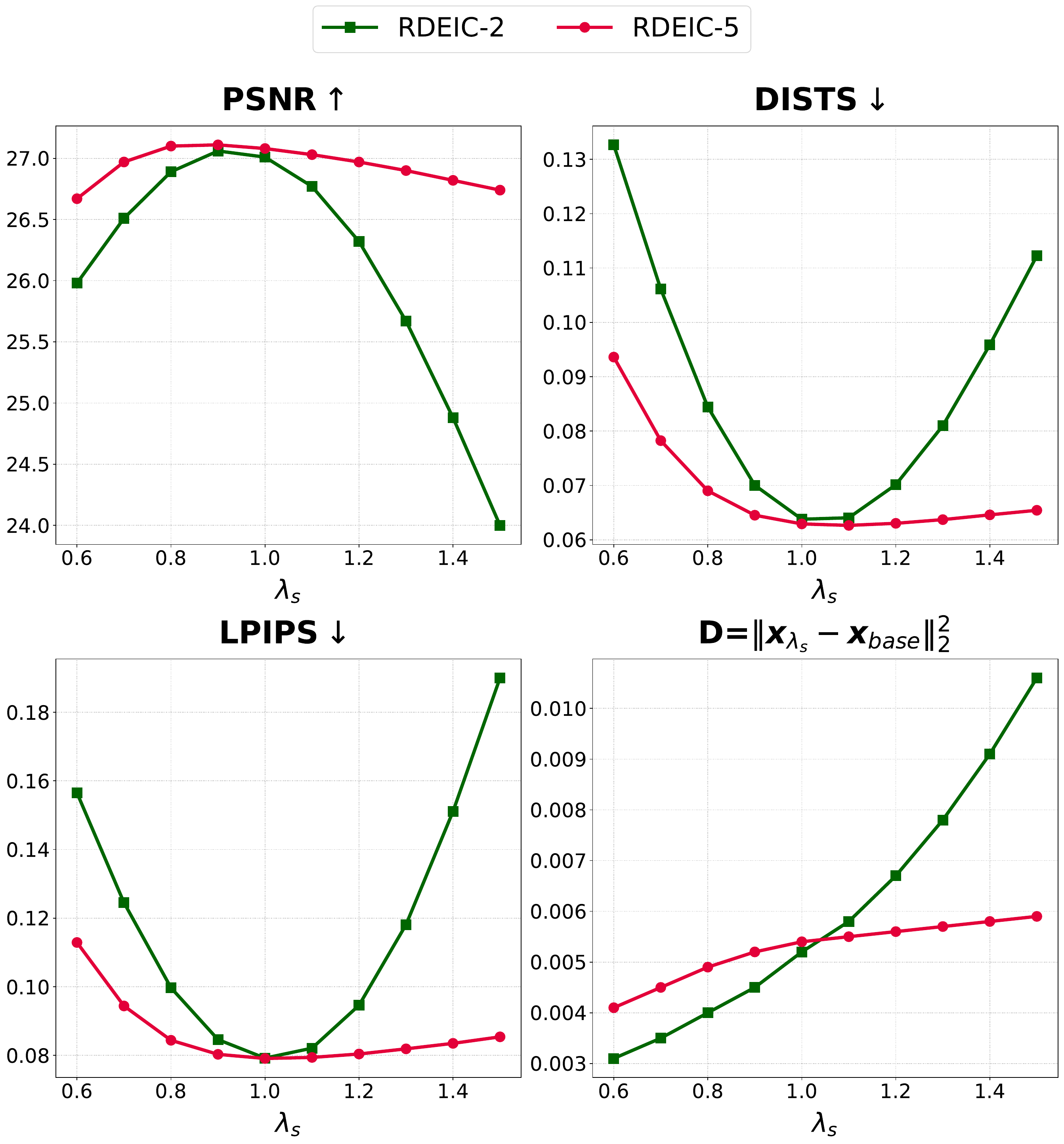}
\caption{{Impact of $\lambda_s$ on quantitative metrics. All metrics are computed on the CLIC2020 dataset. For the computation of $D$, we adopt the reconstruction result of RDEIC-2 at $\lambda_s=0$ (denoted ad $\boldsymbol{x}_{base}$) as the reference image.}}
\label{diversity_q}
\end{figure}

\begin{figure*}[t]\scriptsize
\centering
\includegraphics[width=0.96\textwidth]{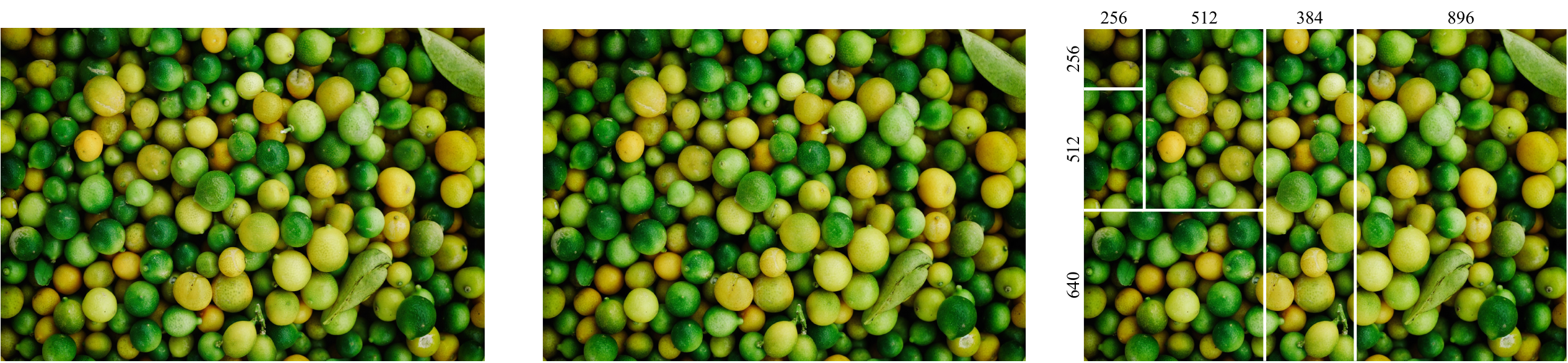}\\
\makebox[0.32\textwidth]{\textbf{(a) Original}}
\makebox[0.32\textwidth]{\textbf{(b) Full-image}}
\makebox[0.32\textwidth]{\textbf{(c) Block-wise}}\\
\vspace{0.1cm}
\makebox[0.32\textwidth]{\textbf{bpp / DISTS$\downarrow$ / MS-SSIM$\uparrow$}}
\makebox[0.32\textwidth]{\textbf{0.0927 / 0.0440 / 0.9352}}
\makebox[0.32\textwidth]{\textbf{0.0930 / 0.0410 / 0.9351}}\\
\caption{{Reconstruction results at different resolutions. (b) the result obtained by directly compressing the entire image, (c) the result obtained by compressing the image patch-by-patch and then stitching the blocks together.}}
\label{multi-resolution}
\end{figure*}

\subsection{Smoothness-sharpness Trade-off}
To fully leverage the generative potential of pre-trained stable diffusion, we introduce a controllable detail generation method that allows users to explore and customize outputs according to their personal preferences. For this experiment, we used the model trained with  $\lambda_r = 1$. The visualization result is shown in Fig. \ref{diversity}. We control the balance between smoothness and sharpness by adjusting the parameter $\lambda_s$, which regulates the amount of high-frequency details introduced into the reconstructed image. Specifically, as the value of $\lambda_s$ increases, the image transitions from a smooth appearance to a progressively sharper and more detailed reconstruction. 

{To provide a more comprehensive analysis of the impact of $\lambda_s$, we further present the corresponding performance curves in Fig. \ref{diversity_q}. As $\lambda_s$ increases, both perceptual quality and reconstruction fidelity initially improve but subsequently decline. To further explain this phenomenon, we adopt the reconstruction result of RDEIC-2 at $\lambda_s=0$ (denoted as $\boldsymbol{x}_{base}$) as the reference image, and compute a distance metric $D$ to quantify the amount of high-frequency content introduced into the reconstructed image $\boldsymbol{x}_{\lambda_s}$:
\begin{equation}
    D = \Vert \boldsymbol{x}_{\lambda_s} - \boldsymbol{x}_{base} \Vert_2^2.
\end{equation}
As shown in Fig. \ref{diversity_q}, the value of $D$ increases monotonically with $\lambda_s$, indicating that more high-frequency information is introduced into the reconstructed image as $\lambda_s$ grows. Initially, introducing a moderate amount of high-frequency detail enhances structural expressiveness and visual realism, resulting in improved perceptual quality and fidelity metrics. Subsequently, when $\lambda_s$ becomes excessively large, the generated high-frequency content may deviate from the original image, resulting in mismatched textures and a subsequent decline in both perceptual quality and distortion performance. Furthermore, the distance curve of RDEIC-5 is flatter than that of RDEIC-2, suggesting that increasing the number of denoising steps improves the model’s robustness to variations in conditional intensity.}

\subsection{{Robustness to Different Image Resolutions}}

{To validate the robustness of our RDEIC across different image resolutions, we conduct experiments on image patches with varying sizes and aspect ratios. Fig. \ref{multi-resolution}(b) presents the result obtained by directly compressing the entire image, while Fig. \ref{multi-resolution}(c) shows the result obtained by compressing the image patch-by-patch and then stitching the blocks together. These visual results demonstrate that our method effectively supports image compression across various dimensions and resolutions, such as $256\times256$, $512\times768$, and $896\times1408$, while maintaining consistent visual quality and compression performance.}

\begin{figure}[t]\scriptsize
\begin{center}
\includegraphics[width=0.48\textwidth]{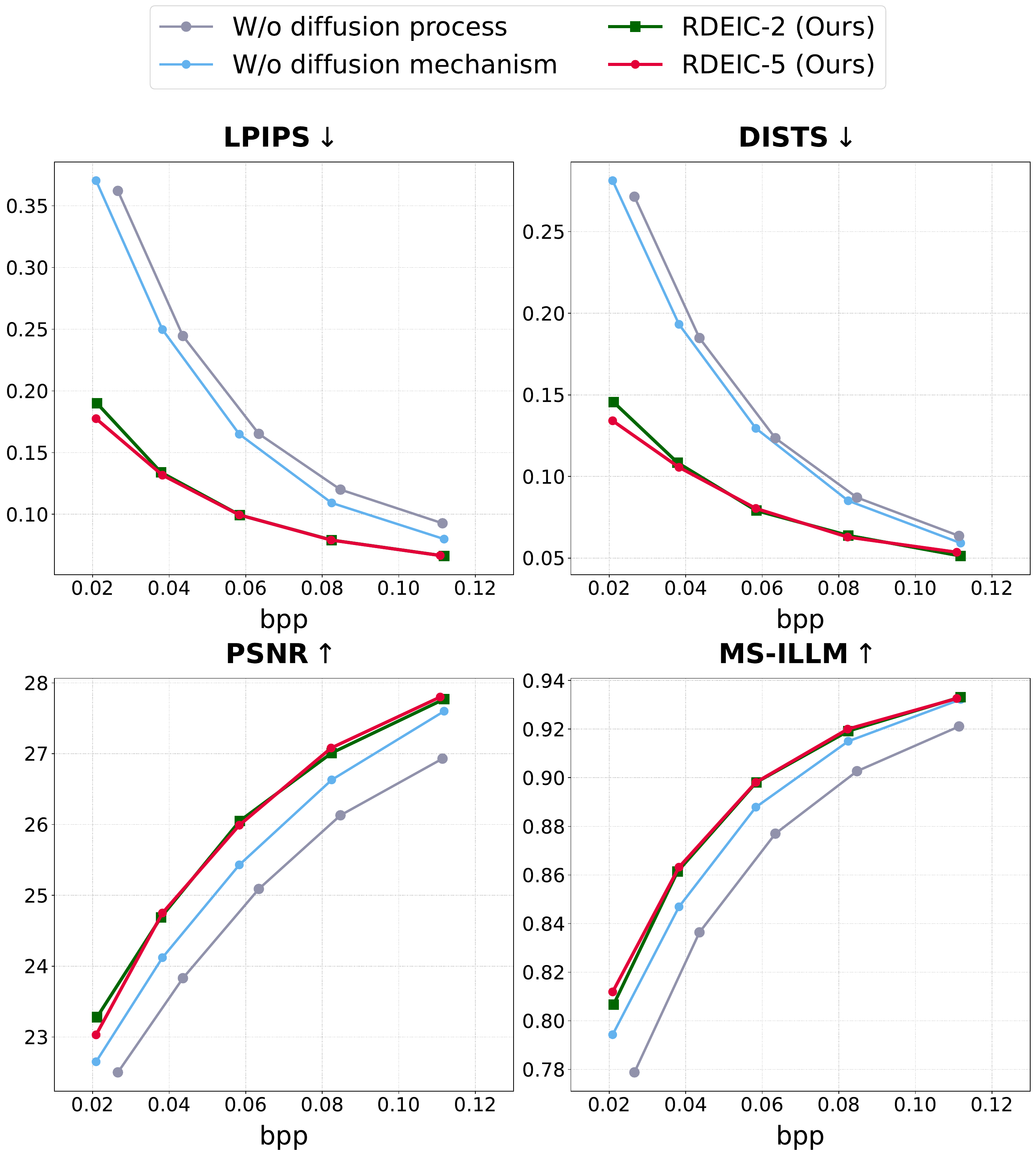}
\end{center}
\caption{Ablation studies of the diffusion mechanism on CLIC2020 dataset. In the \textbf{W/o denoising process} setting, we train the compression module jointly with the noise estimator but bypass the denoising process during inference. In the \textbf{W/o diffusion mechanism} setting, we train the compression module independently, completely excluding the influence of the diffusion mechanism.}
\label{quant_diff}
\end{figure}

\begin{figure*}[t]\scriptsize
\centering
\makebox[0.225\textwidth]{\textbf{Original}}
\makebox[0.225\textwidth]{\textbf{W/o diffusion mechanism}}
\makebox[0.225\textwidth]{\textbf{W/o denoising process}}
\makebox[0.225\textwidth]{\textbf{RDEIC-5}}
\\ \vspace{0.1cm}
\includegraphics[width=0.225\textwidth]{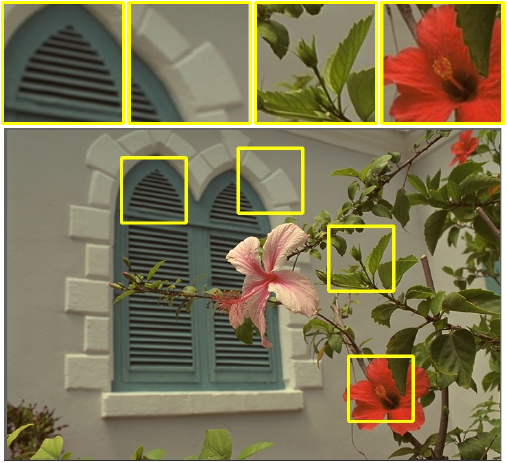}
\includegraphics[width=0.225\textwidth]{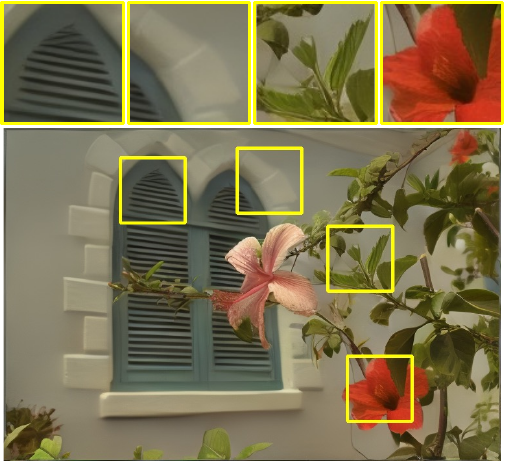}
\includegraphics[width=0.225\textwidth]{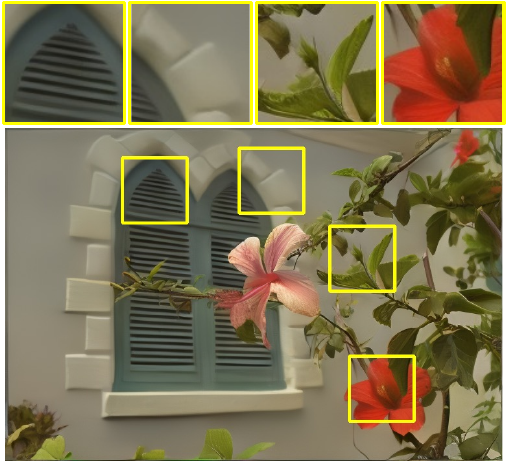}
\includegraphics[width=0.225\textwidth]{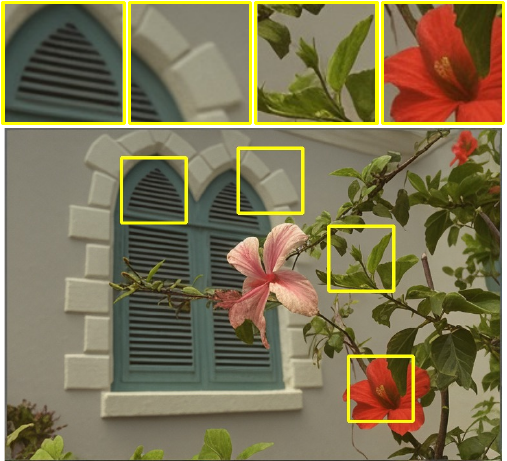}
\\ \vspace{0.1cm}
\makebox[0.225\textwidth]{\textbf{bpp / DISTS$\downarrow$ / PSNR$\uparrow$}}
\makebox[0.225\textwidth]{\textbf{0.0711 / 0.1203 / 24.8603}}
\makebox[0.225\textwidth]{\textbf{0.0672 / 0.1126 / 25.2282}}
\makebox[0.225\textwidth]{\textbf{0.0672 / 0.0709 / 25.9719}}
\caption{Impact of diffusion mechanism on reconstruction results.}
\label{diff_fig}
\end{figure*}

\begin{figure*}[t]\scriptsize
\centering
\makebox[0.265\textwidth]{\textbf{W/o diffusion}}
\makebox[0.265\textwidth]{\textbf{W/ diffusion}}
\makebox[0.17\textwidth]{\textbf{W/o diffusion}}
\makebox[0.17\textwidth]{\textbf{W/ diffusion}}
\\ \vspace{0.1cm}
\includegraphics[width=0.265\textwidth]{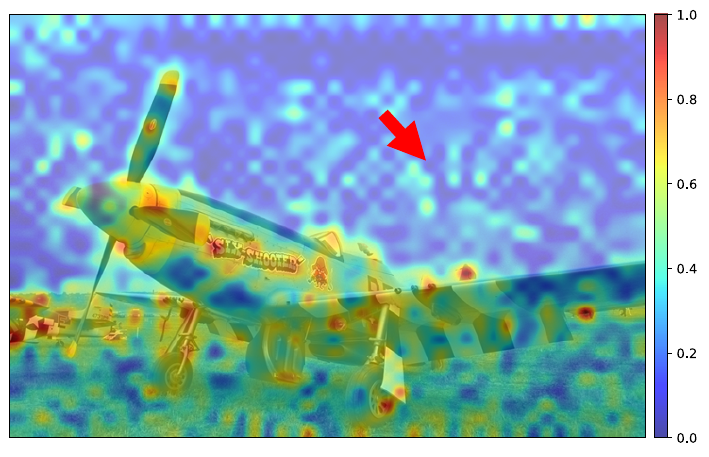}
\includegraphics[width=0.265\textwidth]{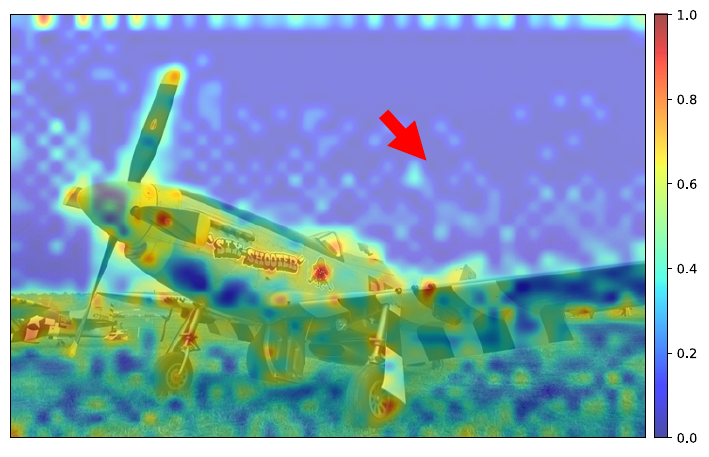}
\includegraphics[width=0.17\textwidth]{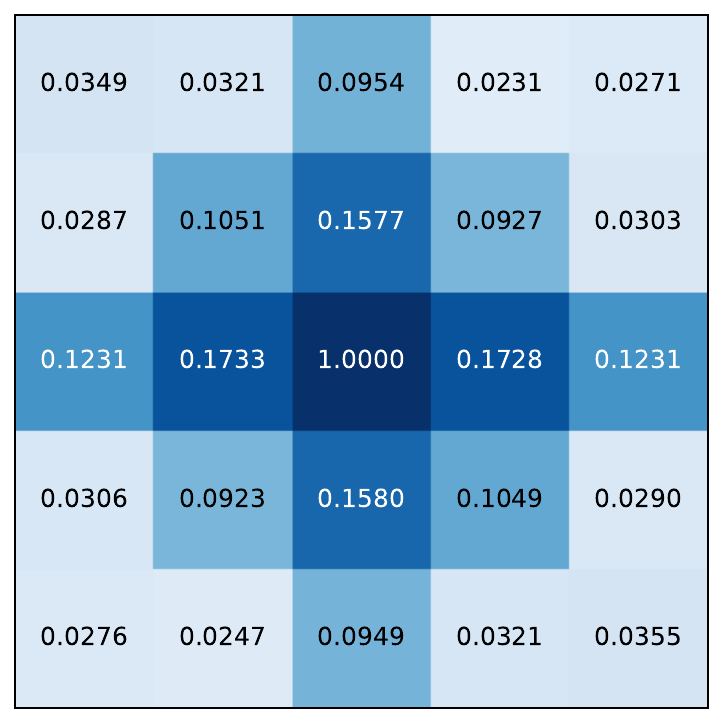}
\includegraphics[width=0.17\textwidth]{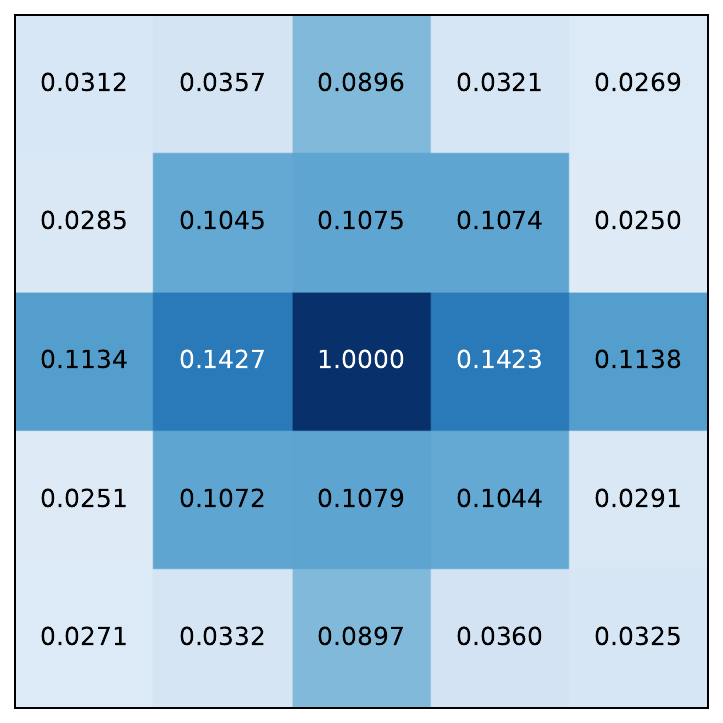}
\makebox[0.265\textwidth]{0.0487 bpp}
\makebox[0.265\textwidth]{0.0460 bpp}
\makebox[0.17\textwidth]{0.0667 bpp}
\makebox[0.17\textwidth]{0.0654 bpp}
\\ \vspace{0.1cm}
\makebox[0.53\textwidth]{\textbf{(a) Bit allocation}}
\makebox[0.34\textwidth]{\textbf{(b) Cross-correlation}}
\caption{Impact of the diffusion mechanism on the compression module. \textbf{W/o diffusion} denotes the compression module trained independently, while \textbf{W/ diffusion} denotes the compression module trained jointly with the noise estimator. All results are obtained from models trained with $\lambda_r=0.5$. (a) An example of bit allocation on the Kodak dataset, with the values normalized for consistency. (b) Latent correlation of $(\boldsymbol{y} - \boldsymbol{\mu}) / \boldsymbol{\sigma}$.}
\label{bit}
\end{figure*}

\begin{figure*}[t]\scriptsize
\begin{center}
\includegraphics[width=0.145\textwidth]{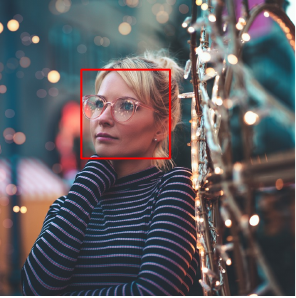}
\includegraphics[width=0.145\textwidth]{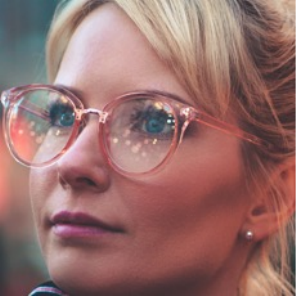}
\includegraphics[width=0.145\textwidth]{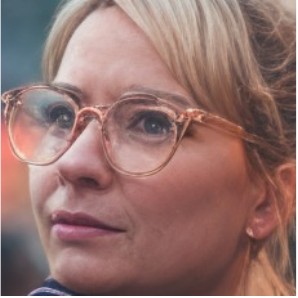}
\includegraphics[width=0.145\textwidth]{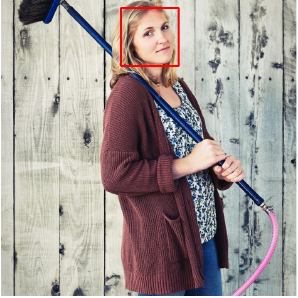}
\includegraphics[width=0.145\textwidth]{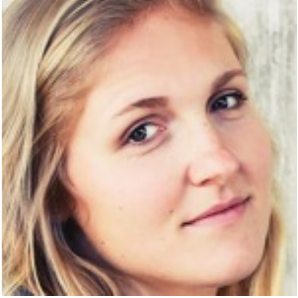}
\includegraphics[width=0.145\textwidth]{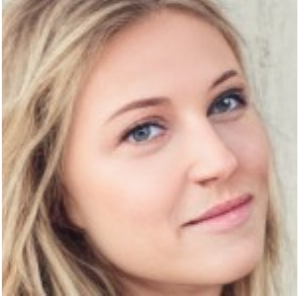}
\\ \vspace{0.1cm}
\makebox[0.148\textwidth]{}
\makebox[0.148\textwidth]{\textbf{Ground Truth}}
\makebox[0.148\textwidth]{\textbf{0.0271 bpp}}
\makebox[0.148\textwidth]{}
\makebox[0.148\textwidth]{\textbf{Ground Truth}}
\makebox[0.148\textwidth]{\textbf{0.0260 bpp}}
\end{center}
\caption{Faces generated at extremely low bitrates.}
\label{face}
\end{figure*}

\subsection{Role of the Diffusion Mechanism}
To further investigate the role of the diffusion mechanism in RDEIC, we design two variants for comparison: 1) \textbf{W/o denoising process}: In this variant, the compression module is trained jointly with the noise estimator, but the denoising process is bypassed during the inference phase. 2) \textbf{W/o diffusion mechanism}: In this variant, the compression module is trained independently, completely excluding the influence of the diffusion mechanism.

As shown in Fig. \ref{quant_diff}, bypassing the denoising process results in significant degradation, particularly in perceptual quality. This demonstrates that the diffusion mechanism plays a crucial role in enhancing perceptual quality during reconstruction. As shown in Fig. \ref{diff_fig}, the diffusion mechanism effectively adds realistic and visually pleasing details. By comparing the performance of W/o diffusion mechanism and W/o denoising process in Fig. \ref{quant_diff} and Fig. \ref{diff_fig}, we observe that the compression module trained jointly with the noise estimator outperforms the one trained independently. This demonstrates that the diffusion mechanism also contributes to the compression module.

Fig. \ref{bit}(a) visualizes an example of bit allocation. It is evident that the model trained jointly with the noise estimator allocates bits more efficiently, assigning fewer bits to flat regions (e.g., the sky in the image). Fig. \ref{bit}(b) visualizes the cross-correlation between each spatial pixel in $(\boldsymbol{y} - \boldsymbol{\mu}) / \boldsymbol{\sigma}$ and its surrounding positions. Specifically, the value at position $(i,j)$ represents cross-correlation between spatial locations $(x,y)$ and $(x+i,y+j)$ along the channel dimension, averaged across all images on Kodak dataset. It is evident that the model trained jointly with the noise estimator exhibits lower latent correlation, suggesting reduced redundancy and more compact feature representations. These results indicate that the diffusion mechanism provides additional guidance for optimizing the compression module during training, enabling it to learn more efficient and compact feature representations.

\subsection{Limitations}
Using pre-trained stable diffusion may generate hallucinated lower-level details at extremely low bitrates. For instance, as shown in Fig. \ref{face}, the generated human faces appear realistic but are inaccurate, which may lead to a misrepresentation of the person’s identity. Furthermore, although the proposed RDEIC has shown promising compression results, the potential of incorporating a text-driven strategy has not yet been explored within our framework. We leave detailed study of this to future work.
\section{Conclusion}
\label{conclusion}
In this paper, we propose an innovative relay residual diffusion-based method (RDEIC) for extreme image compression. Unlike most existing diffusion-based methods that start from pure noise, RDEIC initializes with the compressed latent features of the input image, supplemented with added noise, and reconstructs the image by iteratively removing the noise and reducing the residual between the compressed latent features and the target latent features. To address the discrepancy between the training and inference phases, we propose to fine-tune the model using the entire reconstruction process, further enhancing performance. Inspired by classifier-free guidance (CFG), we introduce a method to balance smoothness and sharpness, overcoming the fixed-step constraint introduced by FSFT. This approach empowers users to explore and customize outputs according to their personal preferences. Extensive experiments have demonstrated the superior performance of RDEIC compared to state-of-the-art methods in terms of both reconstruction quality and computational efficiency.

\bibliographystyle{unsrt}
\bibliography{reference}
\begin{IEEEbiography}[{\includegraphics[width=1in,height=1.25in,clip,keepaspectratio]{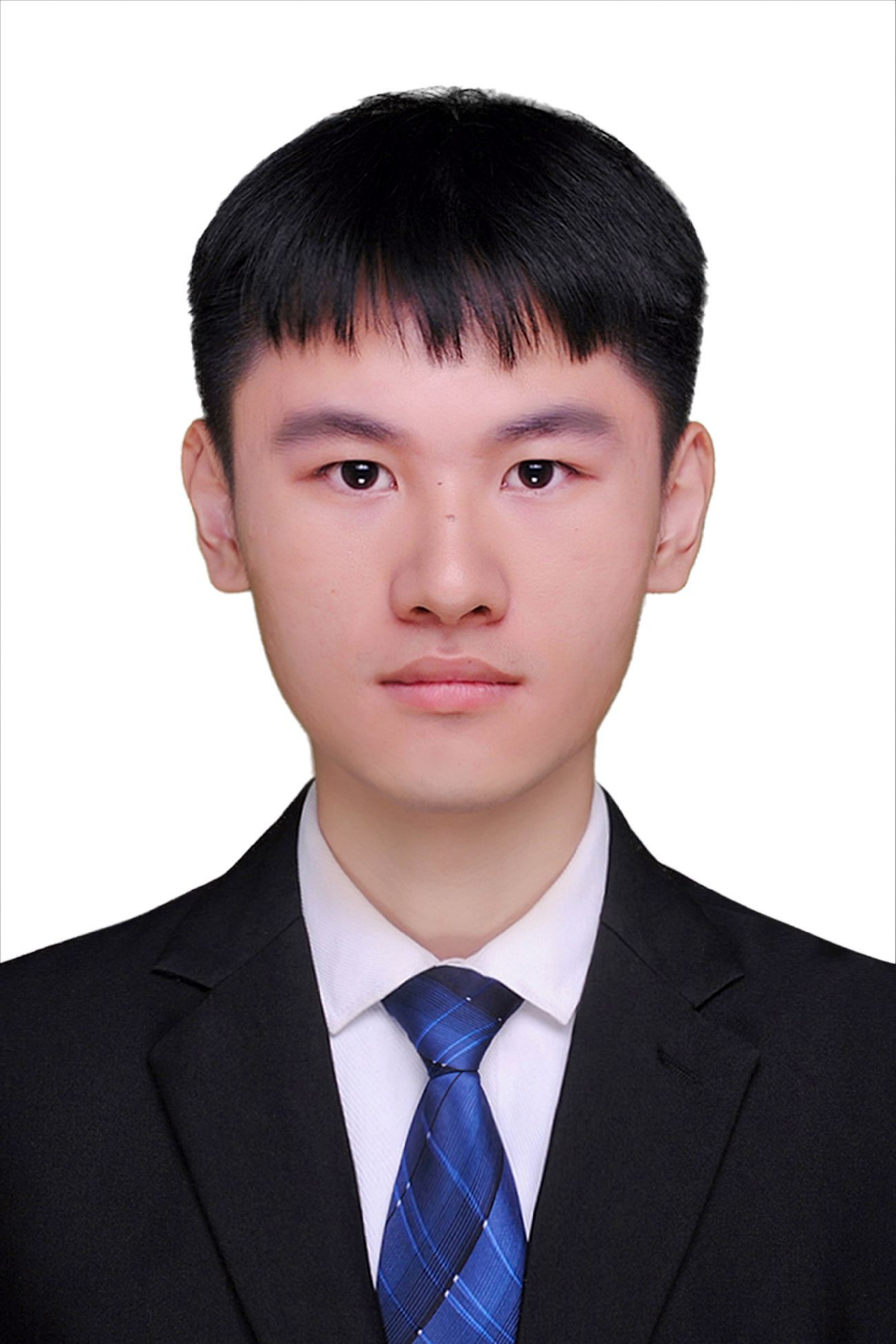}}]{Zhiyuan Li}
is currently pursuing a master's degree with the Institute of Artificial Intelligence and Robotics at Xi’an Jiaotong University. He received his bachelor's degree from Xidian University in 2022. His research interests include image compression, image rescaling, and other visual problems.
\end{IEEEbiography}

\begin{IEEEbiography}
[{\includegraphics[width=1in,height=1.25in,clip,keepaspectratio]{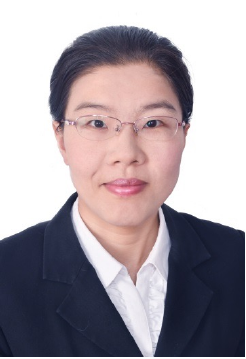}}]{Yanhui Zhou}
received the M. S. and Ph. D. degrees in electrical engineering from the Xi'an Jiaotong University, Xi'an, China, in 2005 and 2011, respectively. She is currently an associate professor with the School of Information and telecommunication Xi’an Jiaotong University. Her current research interests include image/video compression, computer vision and deep learning.
\end{IEEEbiography}

\begin{IEEEbiography}[{\includegraphics[width=1in,height=1.25in,clip,keepaspectratio]{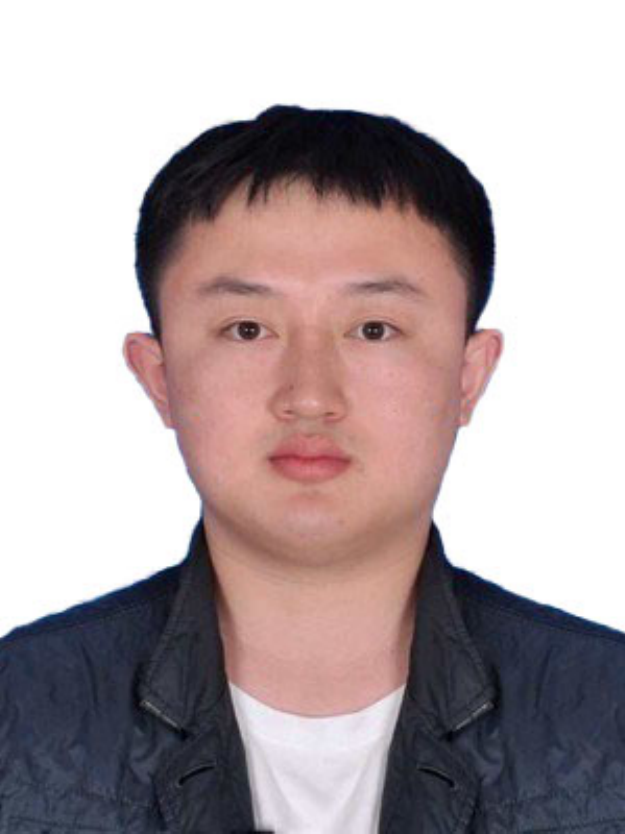}}]{Hao Wei}
is currently a Ph.D. candidate with the Institute of Artificial Intelligence and Robotics at Xi'an Jiaotong University. He received his B.Sc. and M.Sc. degrees from Yangzhou University and Nanjing University of Science and Technology in 2018 and 2021, respectively. His research interests include image deblurring, image compression, and other low-level vision problems.
\end{IEEEbiography}

\begin{IEEEbiography}[{\includegraphics[width=1in,height=1.25in,clip,keepaspectratio]{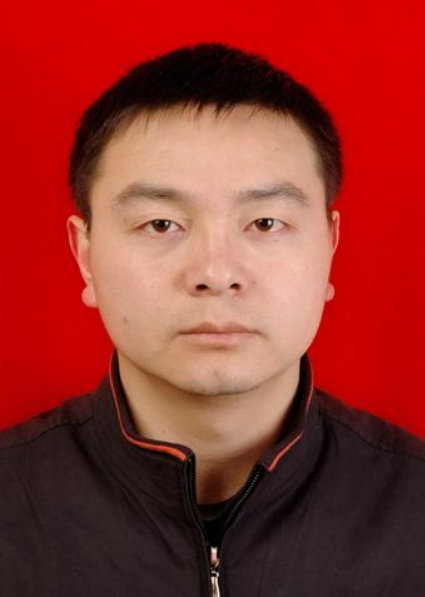}}]{Chenyang Ge}
is currently an associate professor at Xi'an Jiaotong University. He received the B.A., M.S., and Ph.D. degrees at Xi'an Jiaotong University in 1999, 2002, and 2009, respectively. His research interests include computer vision, 3D sensing, new display processing, and SoC design.
\end{IEEEbiography}

\begin{IEEEbiography}[{\includegraphics[width=1in,height=1.25in,clip,keepaspectratio]{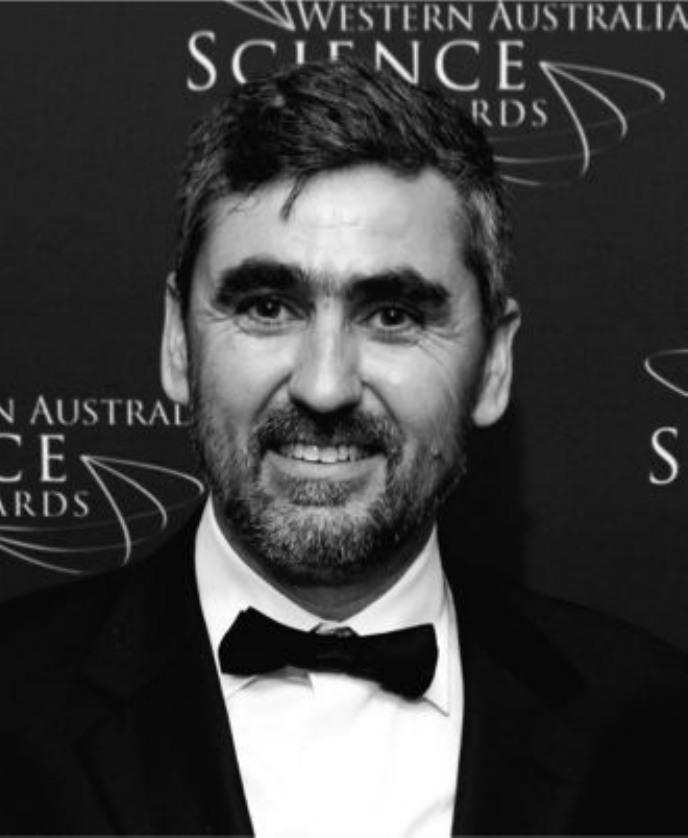}}]{Ajmal Mian}(Senior Member, IEEE)
is a Professor of Computer Science at The University of Western Australia. He is the recipient of three esteemed fellowships from the Australian Research Council (ARC). He has also received several research grants from the ARC, the National Health and Medical Research Council of Australia, US Department of Defense and the Australian Department of Defense with a combined funding of over \$41 million. He received the West Australian Early Career Scientist of the Year 2012 award and the HBF Mid-career Scientist of the Year 2022 award. He has also received several other awards including the Excellence in Research Supervision Award, EH Thompson Award, ASPIRE Professional Development Award, Vice-chancellors Mid- career Award, Outstanding Young Investigator Award, and the Australasian Distinguished Dissertation Award. He is an IAPR Fellow and Distinguished Speaker of the ACM. He also served as a Senior Editor for IEEE Transactions in Neural Networks and Learning Systems and Associate Editor for IEEE Transactions on Image Processing and the Pattern Recognition Journal. He was the General Co-Chair of DICTA 2019 and ACCV 2018. His research interests are in 3D computer vision, machine learning, and video analysis.
\end{IEEEbiography}

\end{document}